\documentclass[11pt]{article}
\pdfoutput=1
\usepackage{amssymb}
\usepackage{psfrag}
\usepackage{amsmath,euscript,array,mathrsfs}
\usepackage{epstopdf}
\newcommand{\beq}{\begin{equation}}
\newcommand{\eeq}{\end{equation}}
\newcommand{\beqs}{\begin{eqnarray}}
\newcommand{\eeqs}{\end{eqnarray}}

\renewcommand{\L}{{\cal L}}

\def\hbar{\hspace{0pt}\raisebox{1pt}{$-$} \hspace{-7pt} h}

\def\r{\rho}

\newcommand{\be}{\begin{equation}}
\newcommand{\ee}{\end{equation}}
\newcommand{\bea}{\begin{eqnarray}}
\newcommand{\eea}{\end{eqnarray}}
\newcommand{\nn}{\nonumber}

\def\co{{\cal O}}

\def\lbldef#1#2{\expandafter\gdef\csname #1\endcsname {#2}}

\def\href#1#2{#2}

\newcommand{\ber}{\begin{eqnarray}}
\newcommand{\eer}{\end{eqnarray}}

\newcommand{\beqar}{\begin{eqnarray}}

\newcommand{\eeqar}{\end{eqnarray}}

\newcommand{\dsl}
  {\kern.06em\hbox{\raise.15ex\hbox{$/$}\kern-.56em\hbox{$\partial$}}}

\newcommand{\eeqarr}{\end{eqnarray}}
\newcommand{\ZZ}{{\rm \kern 0.275em Z \kern -0.92em Z}\;}

\def\CC{{\mathchoice
{\rm C\mkern-8mu\vrule height1.45ex depth-.05ex
width.05em\mkern9mu\kern-.05em}
{\rm C\mkern-8mu\vrule height1.45ex depth-.05ex
width.05em\mkern9mu\kern-.05em}
{\rm C\mkern-8mu\vrule height1ex depth-.07ex
width.035em\mkern9mu\kern-.035em}
{\rm C\mkern-8mu\vrule height.65ex depth-.1ex
width.025em\mkern8mu\kern-.025em}}}

\def\RR{{\rm I\kern-1.6pt {\rm R}}}

\def\ZZ{{\rm Z}\kern-3.8pt {\rm Z} \kern2pt}
\def\IB{\relax{\rm I\kern-.18em B}}
\def\ID{\relax{\rm I\kern-.18em D}}
\def\II{\relax{\rm I\kern-.18em I}}
\def\IP{\relax{\rm I\kern-.18em P}}

\newcommand{\bear}{\begin{eqnarray}}
\newcommand{\eear}{\end{eqnarray}}

\def\to{\rightarrow}

\def\to{\rightarrow}

\def\a{\alpha}

\def\k{\kappa}                    

  \def\w{\omega}
\def\r{\rho}                                     
\def\t{\tau}

\def\lab{\label}
\def\6{\partial}

\def\NO{\nonumber}

\def\bea{\begin{eqnarray}}
\def\eea{\end{eqnarray}}

\def\beqx{\begin{displaymath}}
\def\eeqx{\end{displaymath}}

\newcommand{\bmat}{\left(\begin{array}}
\newcommand{\emat}{\end{array}\right)}

\def\a{\alpha}

\def\k{\kappa}

\def\r{\rho}

\def\t{\tau}

\def\L{\Lambda}

\def\co{{\cal O}}

\def\ra{\rightarrow}                            
\def\bo{{\raise-.3ex\hbox{\large$\Box$}}}               
\def\face{{\raise.2ex\hbox{$\displaystyle \bigodot$}\mskip-2.2mu \llap {$\ddot
        \smile$}}}                                   
\def\>{\rangle}                                      
\def\<{\langle}                                      

\def\leftrightarrowfill{$\mathsurround=0pt \mathord\leftarrow \mkern-6mu
        \cleaders\hbox{$\mkern-2mu \mathord- \mkern-2mu$}\hfill
        \mkern-6mu \mathord\rightarrow$}        
\def\dvec#1{\vbox{\ialign{##\crcr
        \leftrightarrowfill\crcr\noalign{\kern-1pt\nointerlineskip}
        $\hfil\displaystyle{#1}\hfil$\crcr}}}           

\def\-{\hphantom{-}}

\begin{document}

\baselineskip=15.5pt
\pagestyle{plain}
\setcounter{page}{1}

\begin{titlepage}
%

\vspace*{6mm}

\begin{center} 
\Large \bf The Non-SUSY Baryonic Branch: 
Soft Supersymmetry Breaking of ${\cal N}=1$ Gauge Theories
\end{center}

\vskip 10mm
\begin{center}
Stephen~Bennett$^{a,}$\footnote{pystephen@swansea.ac.uk}, 
Elena~Caceres$^{b,c,}$\footnote{elenac@zippy.ph.utexas.edu},
Carlos~N\'u\~nez$^{a,}$\footnote{c.nunez@swansea.ac.uk},
Daniel~Schofield$^{a,}$\footnote{pyschofield@swansea.ac.uk} and
Steve~Young$^{b,}$\footnote{fineline@gmail.com}
 \vskip 4mm

{\it $a$: Department of Physics, Swansea University\\
 Singleton Park, Swansea SA2 8PP, United Kingdom.}
\vskip 5mm

{ \it $b$: Theory Group. Department of Physics\\
University of Texas at Austin, Austin, TX 78727, USA.}

\vskip 5mm

\it{$c$: Facultad de Ciencias, Universidad de 
Colima\\Bernal Diaz del Castillo 340, Colima, Mexico.}

\vspace{0.2in}
\end{center}
\begin{center}
{\bf Abstract}
\end{center}
We study a non-supersymmetric deformation of the field theory dual to the
baryonic branch of Klebanov-Strassler. Using  a combination of analytical (series expansions)  and numerical methods we construct  non-supersymmetric backgrounds that smoothly interpolate between the desired UV and IR behaviors. 
We calculate various observables of the field theory and propose a picture 
of soft breaking by gaugino masses that is consistent with  the various calculations on 
the string side.

\vskip1truecm
\vspace{0.1in}
\end{titlepage}
\setcounter{footnote}{0}

\renewcommand{\theequation}{{\rm\thesection.\arabic{equation}}}

\section{Introduction and Summary}
The Maldacena Conjecture \cite{Maldacena:1997re}
provides what are probably the most effective and controlable
tools to study non-perturbative dynamics of a variety of field theories.
A large  variety  of effects have been discovered or
checked in field theories using a suitable string dual.
Integrability, correlation functions of various interesting
operators (protected or not by symmetries), aspects of 
lower dimensional systems,  applications in
condensed matter and QCD-like systems have been succesfully studied using gauge/gravity duality.

The results above, while more numerous and spectacular
in highly (super)symmetric theories, 
are not restricted to examples of this sort.
As a matter of fact, there are many applications where black holes (and hence dual field theories at finite temperature) play a fundamental role.
In these cases the dynamics is neither driven by SUSY nor by conformal symmetry.

As a result, an interesting problem  is  to construct backgrounds duals to field theories
where supersymmetry has been broken in a  soft way. These systems should
conserve some of the dynamics of the SUSY case with the addition of the
deformations by relevant operators that break the supersymmetry. 
The low energy dynamics should then
be a non-linear superposition of SUSY and non-SUSY effects.
This is an interesting problem, on which it seems feasible to make progress.

In this paper, we will construct duals to field theories in
four dimensions where SUSY has been
explicitly and softly broken by the addition of
relevant operators to the Lagrangian.
The original field theories will be those obtained by a
twisted compactification of five branes wrapping a
calibrated two cycle in the resolved conifold and those obtained by
studying the dynamics of D3 and fractional D5 branes on the tip of a conifold.
Both are non-conformal theories with interesting
low energy dynamics (confinement, R-symmetry breaking, formation
of domain walls, k-strings, etc.)

We will construct our non-SUSY backgrounds 
by finding an explicit solution of the Einstein, 
dilaton and RR-form equations of motion.
We also impose that irrelevant operators are absent from the dynamics
and that the string backgrounds are regular all along the space. 
We will concentrate on the case in which
the SUSY  breaking parameters are small compared the others already
present in the system in the SUSY case.

These will then be examples of backgrounds 
dual to the strongly coupled dynamics of
well understood SUSY field theories in which SUSY has been softly and 
controllably
broken. Some examples of this sort have appeared in the past for 
deformations of well-known SUSY backgrounds, see for example
\cite{Aharony:2002vp}.

Our paper is organized as follows. 
We start in Section \ref{section2} by presenting
the SUSY system. While the fomalism  summarized there does not apply to 
problem of  interest, we do give some details that are  useful
in attempting to construct the non-SUSY 
solutions (in particular large-radius asymptotics).
In Section \ref{sec:susybreaking}, we will the propose a SUSY breaking solution 
in series expansion for large (UV)
and small (IR) values of the radial coordinate. 
We will carefully count the parameters
that control our solutions and find numerical solutions 
interpolating in a smooth way
between the desired UV and IR asymptotics.
Section \ref{seccionnumerica} gives some details of the numerical method. 
In Section \ref{seccionenergy} we
calculate the ADM Energy of the new solutions (with the SUSY solutions 
as reference backgrounds).
In Section \ref{seccionqft} we perform a detailed study of various field theory
quantities, whose strong-coupling result  points us to an interpretation
of the dual field theory being deformed by the 
insertion of relevant operators, like gaugino masses
that break SUSY and may also influence VEVs.
We close with some conclusions and possible interesting problems to be solved
constructing on the results of this paper.
The high technical nature of our work is clear 
from the outline above.
For the benefit  of readers, 
we have  included explicit
technical points in  detailed appendices.

{\it Note Added: While this paper was close to completion, 
we were informed of the
work by Dymarsky and Kuperstein, 
having interesting overlap with 
ours \cite{anatoly}. We thank Anatoly Dymarsky for letting 
us know about this work prior to publication
and discussion on these topics.}

\section{Presentation of the SUSY system}\label{section2}
\setcounter{equation}{0}
In this section we  summarize well established 
aspects of particular
supersymmetric field theories and their 
dual backgrounds. This will be useful
when introducing SUSY breaking deformations.

We start by considering two apparently different field theories.
The first one, we refer to it as `theory I' or `Type I theory' 
(hoping not to cause confusion with the Type I string theory), is a quiver with gauge group 
$SU(n+N_c)\times SU(n)$ and bifundamental matter multiplets $A_i, B_\alpha$ with $i,\a=1,2$.
The  global symmetries are\footnote{The R-symmetry is anomalous, breaking $U(1)_R\to Z_{2N_c}$.}
\beq
SU(2)_L\times SU(2)_R \times U(1)_B \times U(1)_R .
\label{globalsymm}\eeq
These bifundamentals transform under the local and global symmetries as
\beq
A_i=(n+N_c, \bar{n}, 2,1,1,\frac{1}{2}),\;\;\; B_\a=(\bar{n}+\bar{N_c}, 
n,1,2,-1,\frac{1}{2}).
\eeq
There is also a superpotential of the form $
W=\frac{1}{\mu}\epsilon_{ij}\epsilon_{\a\beta}tr[A_iB_\a A_j B_\beta].
$
The field theory is taken to 
be close to a strongly coupled fixed point. In that 
case one can show that the anomalous dimensions should be
$\gamma_{A,B}\sim -\frac{1}{2}$. This field theory is well known 
to be the dual to the Klebanov-Strassler background \cite{KS} and its
generalization to the baryonic branch \cite{BGMPZ}.

The second field theory, that we will call `theory II' 
(again not to be confused with the Type II string!) is obtained 
after a twisted compactification (to four dimensions) 
of six dimensional SUSY $SU(N_c)$ 
Yang-Mills with 16 supercharges. 
This special compactification studied in \cite{MN}, \cite{AD}
preserves four supercharges.
In four dimensional language, the field content 
is a massless vector multiplet
and a `Kaluza-Klein' tower of massive chiral and massive vector multiplets.
The Lagrangian, the weakly coupled mass spectrum and degeneracies
are written in \cite{AD}. The local and global symmetries are 
(the R-symmetry is anomalous, like in the theory~I above), 
\beq
SU(N_c)\times SU(2)_L\times SU(2)_R\times U(1)_R .
\eeq
These two theories, apparently so different, can be connected as discussed in
\cite{MM} and \cite{Elander:2011mh}  via higgsing.
Indeed, giving a particular (classical) baryonic VEV to the fields 
$(A_i, B_\a)$ and expanding around it, the field content and degeneracies
of \cite{AD} is reproduced. 
This weakly coupled field theory connection
has its counterpart in the type IIB solutions dual to each 
of the field theories.
Indeed, it is possible to connect the dual backgrounds 
to field theories I and II, 
using U-duality \cite{MM}. This connection was further studied in
\cite{Elander:2011mh}, \cite{GMNP}, \cite{rotations}, \cite{Caceres:2011zn}. 

We will now explain this connection among the explicit
Type IIB string backgrounds. We start from a background 
describing the strong dynamics of  the `field theory II' 
(the twisted compactification of five branes).
A quite generic configuration of this kind can be compactly 
written using the
$SU(2)$ left-invariant one-forms
\bea\lab{su2}
& & \tilde{\w}_1\,=\, \cos\psi d\tilde\theta\,+\,\sin\psi\sin\tilde\theta
d\tilde\varphi\,\,,\,
\tilde{\w}_2\,=\,-\sin\psi d\tilde\theta\,+\,\cos\psi\sin\tilde\theta
d\tilde\varphi\,\,\nonumber\\
& &\tilde{\w}_3\,=\,d\psi\,+\,\cos\tilde\theta d\tilde\varphi\,\,
\eea
and the vielbeins
\begin{align}
&E^{xi}= e^{\frac{\Phi}{4}} dx_i    ,\;\;\;
E^{\r}=  e^{\frac{\Phi}{4}+k}d\r  ,\;\;\;
E^{\theta}=  e^{\frac{\Phi}{4}+h}d\theta  ,\;\;\;
E^{\varphi}= e^{\frac{\Phi}{4}+h} \sin\theta d\varphi , 
\nn\\
&E^{1}=  \frac{1}{2}e^{\frac{\Phi}{4}+g}(\tilde{\omega}_1 +a d\theta)  ,\;
E^{2}=\frac{1}{2}e^{\frac{\Phi}{4}+g}(\tilde{\omega}_2 
-a\sin\theta d\varphi),										\nn\\
&E^{3}= \frac{1}{2}e^{\frac{\Phi}{4}+k}
(\tilde{\omega}_3 +\cos\theta d\varphi).
\label{vielbeinbefore}
\end{align}
In terms of these, the background and the RR three-form read
\begin{align}
ds_E^2	&= \sum_{i=1}^{10} (E^{i})^2,    \label{f3old}\\
F_3			&= e^{-\frac{3}{4}\Phi}\Big[f_1 E^{123}+ f_2 E^{\theta\varphi 3}
+ f_3(E^{\theta23}+ E^{\varphi 13})+ 
f_4(E^{\r 1\theta}+ E^{\r\varphi 2})   \Big]  \nn
\end{align}
where we defined
\bea
& & E^{ijk..l}=E^{i}\wedge E^{j}\wedge E^{k}\wedge...\wedge E^{l} ,\nonumber\\
& & f_1=-2 N_c e^{-k-2g},\;\;\; f_2= \frac{N_c}{2}e^{-k-2h}(a^2 -
2 a b +1 ),\nonumber\\
& & f_3= N_ce^{-k-h-g}(a-b),\;\;\; f_4=\frac{N_c}{2}e^{-k-h-g}b' .
\eea
The dilaton, as usual, is a function of the radial coordinate $\Phi(\r)$
and we have set $\alpha^{\prime}g_s=1$.

The full background is then determined by 
solving the equations of motion for the 
functions $(a,b,\Phi,g,h,k)$. 
A system of BPS equations is derived using 
this Ansatz (see appendix of
reference \cite{Casero:2006pt}). These non-linear and coupled 
first order equations
can be arranged in a convenient form, by rewriting the
functions of the background in terms of a new basis
of functions $P(\rho),\
Q(\rho),Y(\rho),\ \tau(\rho),\ \sigma(\rho)$ that decouples the equations (as explained in
\cite{HoyosBadajoz:2008fw}-\cite{Casero:2007jj}). 
We quote this change of basis
in our Appendix \ref{backgroundfunctionsappendix}.

Using these new variables, one can manipulate 
the decoupled BPS equations, solving most of them
and obtaining a
single decoupled second order equation for $P(\rho)$. 
All other functions are
obtained from $P(\rho)$ --- see \cite{HoyosBadajoz:2008fw} 
and our Appendix \ref{backgroundfunctionsappendix} for details.
The second order equation mentioned above  reads
\beq
P'' + P'\Big(\frac{P'+Q'}{P-Q} +\frac{P'-Q'}{P+Q} - 4 
\coth(2\rho-2{\rho}_0)
\Big)=0.
\label{master}
\eeq
We will refer to Eq.~(\ref{master}) 
as the {\it master equation}: this is the only 
equation that needs solving in order to 
generate the large classes of solutions of Type IIB dual to
``field theory II'' 
in different circumstances 
(vacua, insertion of operators in the Lagrangian, etc.)
\footnote{As an example, the solution $P=2N_c \r$ 
gives the background of \cite{MN},
\cite{Chamseddine:1997nm}. This solution and those with the same large $\r$
asymptotics
will not be the focus 
of this paper.}.

In this paper, we will not be concerned with 
SUSY solutions, but they will play an 
important guiding role. 
We summarize below the small and large $\r$
expansions of the function $P(\r)$.
\subsection{Aspects of the SUSY solutions}
Let us start from the solution of the master equation (\ref{master})
for large values of the radial coordinate (describing 
the UV of the field theory II).
The SUSY solutions  
have an expansion for 
$\r\to \infty$ of the form,
\bea
& & P=e^{4\rho/3}\biggl[ c_+  
+\frac{e^{-8\r/3} N_c^2}{c_+}\left(
4\r^2 - 4\r +\frac{13}{4} \right)+ e^{-4\r}\left(
c_- -\frac{8c_+}{3}\r \right)\nonumber\\
& &  + \frac{N_c^4 e^{-16\r/3}}{c_+^3}
\left(\frac{18567}{512}-\frac{2781}{32}\r +\frac{27}{4}\r^2 -36\r^3\right)  \biggr]
\label{UV-II-N}
\eea
Notice that this expansion 
involves two integration constants, $c_+>0$ and $c_-$. 
The background functions
at large $\r$ are written
in Appendix \ref{backgroundfunctionsappendix}.

Regarding the IR expansion, 
we look for solutions with $P\to 0$ 
as $\r\to 0$, 
in which case we find
\be
P= h_1 \r+ \frac{4 h_1}{15}\left(1-\frac{4 N_c^2}{h_1^2}\right)\r^3
+\frac{16 h_1}{525}\left(1-\frac{4N_c^2}{3h_1^2}-
\frac{32N_c^4}{3h_1^4}\right)\r^5+\co(\r^7),
\label{P-IR}\ee
where $h_1$ is again an arbitrary constant, 
there is of course another integration constant, taken to zero here, to avoid singularities. 
This gives background functions
that are quoted in Appendix \ref{backgroundfunctionsappendix}.
Of course, there is a smooth numerical interpolation between both 
expansions. However, there is then only one independent parameter; given a value for one of $\{c_+, c_-, h_1\}$, the requirement that the solution matches \emph{both} expansions is sufficient to determine the values of the other two.

As explained in \cite{Elander:2011mh}, this solution corresponds to
a dual field theory II in the presence of a dimension-eight 
operator inserted in the Lagrangian which ultimately couples 
the field theory to gravity.
This calls for a completion in the context of field theory.
This is achieved with  the U-duality of \cite{MM} (which we will 
sometimes refer to as 
the `rotation').

After the U-duality described in \cite{MM} is applied, we
define the new vielbein 
(which we use in the following),
\begin{align}
	e^{xi}			&= e^{\frac{\Phi}{4}}\hat{h}^{-\frac{1}{4}} dx_i    ,\;\;\;
	e^{\r}	 		=  e^{\frac{\Phi}{4}+k} \hat{h}^{\frac{1}{4}}d\r  ,\;\;\;
	e^{\theta}	=  e^{\frac{\Phi}{4}+h} \hat{h}^{\frac{1}{4}}    d\theta  ,\;\;\;
	 e^{\varphi}= e^{\frac{\Phi}{4}+h} \hat{h}^{\frac{1}{4}} 	\sin\theta d\varphi    ,\nonumber\\
	e^{1}				&=  \frac{1}{2}e^{\frac{\Phi}{4}+g} \hat{h}^{\frac{1}{4}}	(\tilde{\omega}_1 +a d\theta)  ,\;\;\;
	e^{2}				=		\frac{1}{2}e^{\frac{\Phi}{4}+g} \hat{h}^{\frac{1}{4}}	(\tilde{\omega}_2 	-a\sin\theta d\varphi)   ,\nonumber\\ 
	e^{3}				&= \frac{1}{2}e^{\frac{\Phi}{4}+k} \hat{h}^{\frac{1}{4}}	(\tilde{\omega}_3 +\cos\theta d\varphi).
	\label{vielbeinafter}
\end{align}
The newly generated  metric, RR and NS fields are
\bea
& & ds_E^2= \sum_{i=1}^{10} (e^{i})^2,\nonumber\\
& & F_3= \frac{e^{-\frac{3}{4}\Phi}}{\hat{h}^{3/4}}
\Big[f_1 e^{123}+ f_2 e^{\theta\varphi 3}
+ f_3(e^{\theta23}+ e^{\varphi 13})+ 
f_4(e^{\r 1\theta}+ e^{\r\varphi 2})   \Big]\nonumber\\
& & H_3=-\k \frac{e^{\frac{5}{4}\Phi}}{\hat{h}^{3/4}}
\Big[-f_1 e^{\theta\varphi \r} - f_2 e^{\r 12}
- f_3(e^{\theta 2\r}+ e^{\varphi 1\r})+
f_4(e^{ 1\theta 3}+ e^{\varphi 2 3})   \Big]\nonumber\\
& & C_4= -\k \frac{e^{2\Phi}}{\hat{h}} 
dt\wedge dx_1 \wedge dx_2\wedge dx_3,  \nonumber\\
& & F_5= \k e^{-\frac{5}{4}\Phi -k}\hat{h}^{\frac{3}{4}}
\partial_\r \left(\frac{e^{2\Phi}}{\hat{h}}\right)
\Big[e^{\theta\varphi 123 }- e^{tx_1 x_2 x_3 \r}   \Big].
\label{configurationfinal}
\eea
We have defined
\beq
\hat{h}=1-\k^2 e^{2\Phi},
\eeq
where $\k$ is a constant that we will choose to be $\k= e^{-\Phi(\infty)}$,
forcing the dilaton to be 
bounded at large distances.
The rationale for this choice is to obtain 
a dual QFT decoupled from gravity. 
Details of this were
carefully discussed in \cite{Elander:2011mh}, \cite{Caceres:2011zn}.
The tuning $\k= e^{-\Phi(\infty)}$ (also chosen in \cite{Dymarsky:2005xt}, 
though in slightly different notation) is the geometric version
of the fact that, in order 
to eliminate an irrelevant operator in the dual field theory I, 
we have to finely-tune the matter content and 
the gauge group with which we will UV-complete the 
theory II after un-higgsing from the single node to the quiver.
See \cite{Elander:2011mh} for a complete explanation.
We will now move to study SUSY breaking deformations.

\section{The SUSY-breaking deformation}\label{sec:susybreaking}
\setcounter{equation}{0}
The goal is to find a non supersymmetric solution with the 
same symmetries and structure as the ones described above.
We proceed as follows: we will find a non-SUSY generalization
of the system in eq.(\ref{f3old}). We will solve the 
equations corresponding to Einstein, Maxwell, dilaton and Bianchi equations 
of the system. The nice properties of
the SUSY formalism just explained do not apply. We will 
then propose series expansions for the individual background functions
$\Phi,h,g,k,a,b$. With the experience gained in the SUSY example,
especially keeping in mind the expansions quoted in 
eqs. (\ref{gggg}, \ref{SUSYIREXP}), we propose similar asymptotics.
\subsection{Asymptotic expansions}
In the UV (large values of $\r$) 
our expansions take the form,
\bea
& & e^{2h}\sim \sum_{i=0}^\infty \sum_{j=0}^{i}H_{ij} \r^j e^{4(1-i)\r/3},
\;\; \frac{e^{2g}}{4} \sim \sum_{i=0}^\infty \sum_{j=0}^{i}G_{ij} 
\r^j e^{4(1-i)\r/3},\nonumber\\
& & \frac{e^{2k}}{4}\sim \sum_{i=0}^\infty \sum_{j=0}^{i} K_{ij} 
\r^j e^{4(1-i)\r/3},\;\;
e^{4\Phi}\sim \sum_{i=1}^\infty \sum_{j=0}^{i}\Phi_{ij} 
\r^j e^{4(1-i)\r/3},\nonumber\\
& & b(\r)\sim \sum_{i=1}^\infty \sum_{j=0}^{i}V_{ij} \r^j e^{2(1-i)\r/3},\;
a(\r)\sim \sum_{i=1}^\infty \sum_{j=0}^{i}W_{ij} \r^j e^{2(1-i)\r/3}.
\label{UVexpansionnonsusy}
\eea
We have found that a generic solution of this sort 
can be written in terms of nine integration constants.
These constants are free; all other coefficients in 
the series expansion can be written in terms of them. 
The independent constants are taken to be,
\bea
K_{00},\; K_{30},\; H_{10}, H_{11},\Phi_{10}, \Phi_{30}, W_{20}, W_{40}, V_{40}.
\label{constzzzz}
\eea
Note that we have found 
the constants $V_{21}, W_{21}$ must 
vanish for this to be a solution. 
Also, we imposed that terms that would 
spoil the UV behavior of the SUSY solution
(corresponding to irrelevant operators in the dual QFT) 
are absent from our expansions.

Without loss of generality, 
we relabel the UV parameters in eq.(\ref{constzzzz}) 
to make contact with the SUSY case 
(see appendix \ref{backgroundfunctionsappendix}):
\begin{align}
	W_{40} 		= 2e^{\rho_o},	\qquad
	K_{00} 		= \frac{2c_+}{3},	\qquad
	\Phi_{10} = e^{4\Phi_\infty},	\nn\\
	H_{10}		=	\frac{Q_o}{4}, \qquad
	K_{30}		=	\frac{c_- -64e^{4\rho_o}c_+^3}{48c_+^2}.
\end{align}
The independent parameters are then
\begin{align}
c_+,\; c_-,\;\Phi_\infty,\; Q_o,\; \rho_o,\;  H_{11},\; W_{20},\; \Phi_{30},\;  V_{40},
\label{UVparam}
\end{align}
and we can recover the SUSY case by setting
\begin{align}
	H_{11}=\frac12	,\;
	W_{20}=0				,\;
	\Phi_{30}= - \frac{3e^{4\Phi_\infty}}{4c_+^2} (3+4Q_o)		,\;
	V_{40}=2 e^{2\rho_o} (1+Q_o).
\label{recoverSUSYUV}
\end{align}

For small values of the radial coordinate
(which we take to end at $\r=0$) we will propose an expansion
of the form (again, imposing regularity of the solution),
\bea
& & e^{2h}\sim \sum_{j=2}^{\infty}h_{j} \r^j,\;\;\;
\frac{e^{2g}}{4}\sim \sum_{j=0}^{\infty}g_{j} \r^j,
\;\;\;\frac{e^{2k}}{4}\sim \sum_{j=0}^{\infty}k_{j} \r^j,\\
& & e^{4\Phi}\sim \sum_{j=0}^{\infty}f_j \r^j,\;\;\;
a(\r)\sim \sum_{j=0}^{\infty}w_j \r^j,\;\;\;
b(\r)\sim \sum_{j=0}^{\infty}v_j \r^j.\nonumber
\label{IRnonsusyexpansion}
\eea
In this case the free parameters are $k_0$, $f_0$, $k_2$, $v_2$ and $w_2$.
For any value of these numbers we find a 
solution. To make contact with the SUSY solution 
in eq.(\ref{SUSYIREXP}), we relabel $k_0=h_1/2$ and $f_0=e^{4\phi_0}$. We then 
recover the SUSY solution if the remaining three parameters take the values
\begin{align}
	k_2 = \frac{2}{5h_1}(h_1^2-4)	,\quad
	v_2	=	-\frac23								,\quad
	w_2	=	\frac{8}{3h_1}-2.
\end{align}

\label{parametercounting}
If we want to restrict our attention to those solutions which match both the UV \emph{and} IR expansions described above, we expect to have fewer independent parameters. To see this, note that we can describe the solutions either by the IR boundary conditions, so that they are parameterised by the five IR parameters $\{h_1,\phi_0,k_2,v_2,w_2\}$, or by the UV boundary conditions, giving a parameterisation in terms of 
\begin{align}
\{c_+,c_-,\Phi_\infty,Q_o, \rho_o,H_{11},W_{20},\Phi_{30},V_{40}\}. \nn	
\end{align}
If a solution exists which connects our IR and UV expansions, the functions resulting from these two parameterizations must be the same. We can formally express this as a system of twelve equations\footnote{We write the functions resulting from a given choice $\{h_1,\phi_0,k_2,v_2,w_2\}$ of the IR parameters in the form $g_{h_1,\phi_0,k_2,v_2,w_2}(\rho)$. Similarly, the expressions 
of the form $g_{c_+,c_-,\Phi_\infty,Q_o, \rho_o,H_{11},W_{20},\Phi_{30},V_{40}}(\rho)	$ refer to the functions resulting from a given choice of the UV parameters. },
\begin{center}
\begin{tabular}{r @{\ } c @{\ } l		@{\qquad} 	 	r @{\ } c @{\ } l}
	$		g_{h_1 \dots w_2}(\rho)		$	&	$=			$	&	$g_{c_+ \dots  V_{40}}(\rho)	$,	&				
	$		g_{h_1 \dots w_2}'(\rho)	$	&	$=			$	&	$g_{c_+ \dots  V_{40}}'(\rho)	$,	\\
	$		h_{h_1 \dots w_2}(\rho)		$	&	$=			$	&	$h_{c_+ \dots  V_{40}}(\rho)	$,	&				
	$		h_{h_1 \dots w_2}'(\rho)	$	&	$=			$	&	$h_{c_+ \dots  V_{40}}'(\rho)	$,	\\
																	&	$\vdots	$	&																 		&				
																	&	$\vdots	$	&																 		\\
	$		b_{h_1 \dots w_2}(\rho)		$	&	$=			$	&	$b_{c_+ \dots  V_{40}}(\rho)	$,	&				
	$		b_{h_1 \dots w_2}'(\rho)	$	&	$=			$	&	$b_{c_+ \dots  V_{40}}'(\rho)	$.
\end{tabular}
\end{center}

However, the derivative of one of the functions can 
be expressed in terms of the other derivatives and the 
functions themselves using the constraint, so only 
eleven of these equations can be independent. We therefore expect to 
be able to solve for eleven of the fourteen parameters, leaving only three independent.

We know that the dilaton can be shifted without 
otherwise affecting the solution, so one of these parameters must be 
either $\phi_0$ or $\Phi_\infty$. Additionally, we know that we also need 
one of $h_1$, $c_+$ or $c_-$ to describe the class of SUSY solutions. 
The third parameter therefore breaks SUSY.

Note that there does not appear to be anything to stop us choosing, 
say, $Q_o$ to parameterise the SUSY-breaking, 
despite the fact that UV expansions with $Q_o\ne -N_c$ do not break SUSY. 
This is explained by the fact that SUSY solutions with $Q_o\ne-N_c$ 
do not have a regular IR. 
If we simultaneously demand that $Q_o\ne-N_c$ \emph{and} 
the IR is of the form (\ref{IRexpg}--\ref{IRexpb}) we must therefore have a non-SUSY solution. 
However, it seems conceptually simpler to choose the third parameter 
to be one which \emph{explicitly} breaks SUSY. For our three independent parameters, 
we might select (in the IR) $\{h_1,\phi_0,w_2\}$, or (in the UV) 
$\{c_+, \Phi_\infty, W_{20}\}$.

In the next section, we will study the challenging 
numerical problem of 
finding a solution that interpolates between the small 
and the large $\r$ expansions in 
eqs.(\ref{UVexpansionnonsusy})-(\ref{IRnonsusyexpansion}).

To summarize: we want to 
find a 
numerical solution for the functions above. This will provide us with
a  non-SUSY deformation of the background in eq.(\ref{f3old}) 
dual to field theory of type II. Applying the U-duality in 
\cite{MM}-\cite{Caceres:2011zn}
we
construct a non-SUSY background of the form given 
in eq.(\ref{configurationfinal}), 
dual to a non-SUSY version of the field theory I.

\section{Numerical analysis}\label{seccionnumerica}
\setcounter{equation}{0}
In this section we show that for some values of the free constants in the IR and in 
the UV we can connect the asymptotics numerically. We first briefly describe 
our method of relating the IR and UV parameters, the details of which 
are relegated to  appendix \ref{NumApp}, before presenting a sample solution 
which smoothly connects the IR and UV asymptotics of the previous section.

Our approach is to solve the equations of motion (\ref{eq:eomg}--\ref{eq:eomb}) 
numerically, using the expansions (see appendix \ref{explicitexpansionsappendix}) 
as boundary conditions. We start from the IR expansions (\ref{IRexpg}--\ref{IRexpb}), 
meaning that our numerical 
solutions are described by the three SUSY-breaking parameters 
$\{k_2,v_2,w_2\}$, 
in addition to those already present in the SUSY case $h_1$ and $\phi_0$.

However, we have seen in section \ref{parametercounting} 
that we expect only three independent parameters in total, one of which breaks SUSY. 
This would mean that in the non-SUSY case we cannot treat even 
the five IR parameters as independent. This is confirmed by our numerical analysis: 
a generic choice of the IR parameters yields UV behaviour of 
the form $b\sim \pm e^{2\rho}$ and $e^{2g}\sim e^{2h}\sim e^{2k} \sim e^{8\rho/3}$, 
which is incompatible with our expansions \eqref{UVexpansionnonsusy}. 
To obtain a solution connecting the IR and UV expansions, then, we have 
to determine both an appropriate combination of values for the IR parameters
and the corresponding values for the UV parameters.

We achieve this using the method described in detail in appendix \ref{NumApp}. In outline, we start with a manual search of the IR parameter space, using Mathematica's {\ttfamily NDSolve} to obtain numerical solutions. Having obtained a solution with the desired UV behaviour ($b\sim e^{-2\r/3}$, $g\sim h\sim k\sim e^{4\rho/3}$) we optimise the match to the UV expansions (\ref{UVexpg}--\ref{UVexpb}) with respect to the parameters \eqref{UVparam} using {\ttfamily NMinimize}.

\begin{figure}[htbp]
	\centering
		\includegraphics{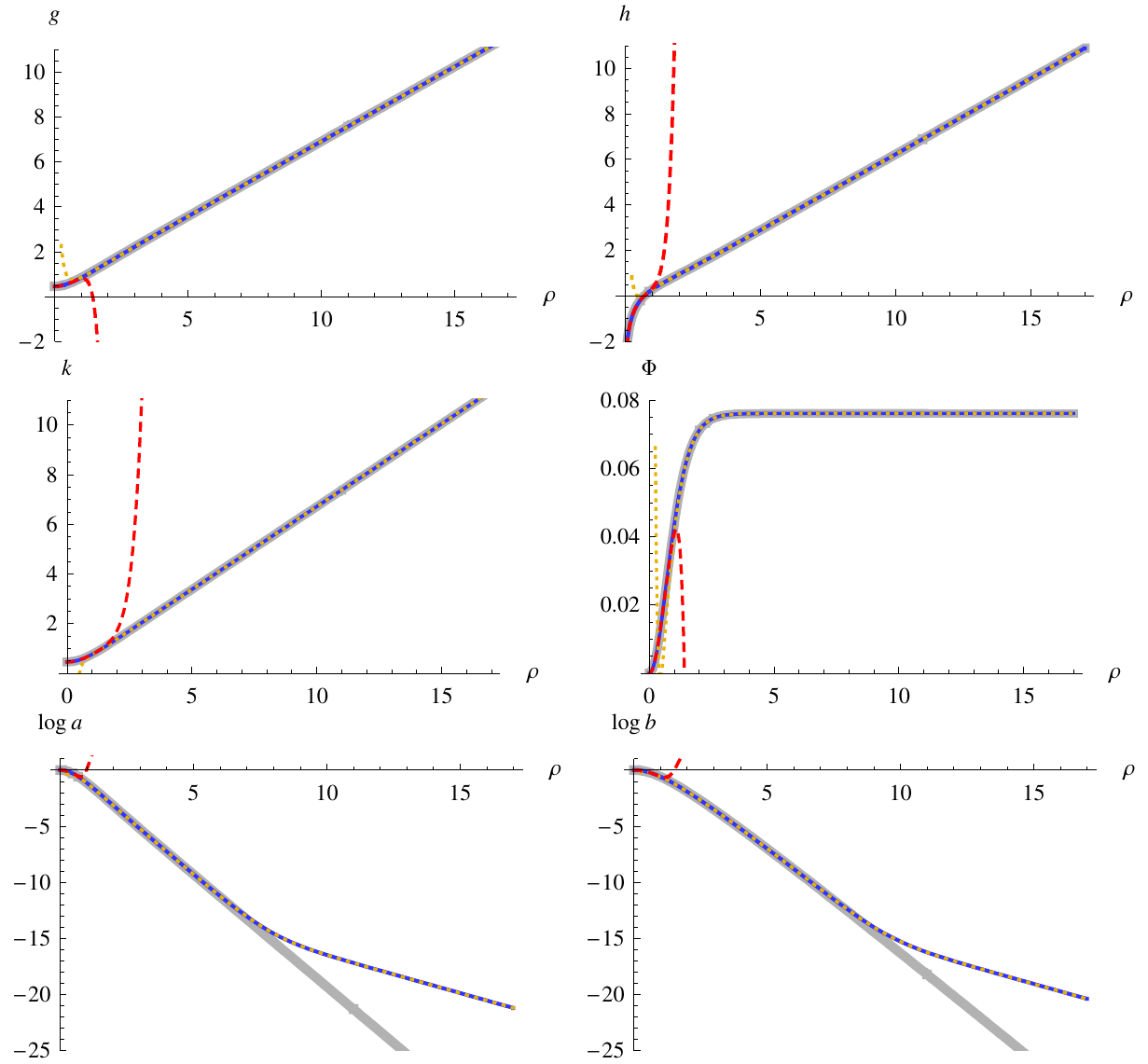}
	\caption{Plots of the functions $g$, $h$, $k$, $\Phi$, $\log a$ and $\log b$, obtained numerically (solid blue), together with the IR (dotted red) and UV (dashed orange) expansions (appendix \ref{explicitexpansionsappendix}), with small deviations from the SUSY values of the parameters. The SUSY solution (grey) is included for comparison.	
	}
	\label{fig:functionPlots}
\end{figure}

An example of solution is shown in figure \ref{fig:functionPlots}. 
As is expected from the expansions (\ref{UVexpg}--\ref{UVexpb}), 
the most significant modification with respect to the SUSY solution 
is the presence of the $e^{-2\rho/3}$ behaviour in the UV of $a$ and $b$. 
The size of this effect is controlled by the SUSY-breaking parameter $W_{20}$; 
in the SUSY case we have $W_{20}=0$, resulting in $a\sim e^{-2\rho}$ 
and $b\sim \rho e^{-2\rho}$. The other functions are modified 
at higher order in (\ref{UVexpg}--\ref{UVexpb}), and as expected no effect 
is visible in figure \ref{fig:functionPlots}. 
See Appendix \ref{appendixdetailsnum}
for details of the numerical analysis.

\section{Energy }\label{seccionenergy}
\setcounter{equation}{0}

In this section we study the energy of the non-SUSY solutions found above.
For any stationary spacetime admitting 
foliations by a spacelike hypersurface $\Sigma_t$, 
the free energy and the energy are 
related via the thermodynamic relation $ F= E - T S $. 
Here we are considering  $T=0$ backgrounds
and so we expect 
$F=E$. In this section  we will first calculate the ADM energy $E$, 
for the solutions before the U-duality --- we will refer to the U-duality
of reference \cite{MM} as `rotation'. We will 
then repeat this calculation for the solutions after 
rotation and show that the energies 
before and after rotation  are  equal. 
As a check of our results, in Appendix \ref{Freeenergyappendix} we  obtain  
the free energy using the on-shell action method  and show that $F=E$.

\subsection{ADM energy}
Consider a non-asymptotically flat 10-dimensional background. 
Let $\Sigma_t$ be a 9-dimensional constant-time 
slice whose 8 dimensional boundary is a 
constant-radius surface $S_t^\infty$. 
The regularized internal energy E  is   defined as \cite{Hawking:1995fd},
\begin{align}\label{eq:hawking-horowitz}
	E&= -\frac{1}{8\pi} \int_{S_t^\infty}\left[ N_t\left( ^8K - ^8K_0\right) + N^\mu_t p_{\mu \nu} n^\nu\right]d S_t^\infty .
\end{align}
$N_t$ is the lapse function, $N^\mu_t$ is the shift vector, $p_{\mu\nu}$ the momentum conjugate to the time derivative in the constant time-slice, $^8K$ and  $^8K_0$ are  the 
extrinsic curvatures of  the 8 dimensional boundary $S_t^\infty$, for the background under consideration and the reference background respectively. Finally $n^\nu$ is the spatial unit vector normal to the constant radius-surface $S_t^\infty$.
It is required that both geometries induce the same metric on $S_t{^\infty}$.
The matter fields should also agree at $S_t^\infty$ or at least the difference should tend to zero as  $S_t^\infty$  goes to infinity.
We will choose a SUSY background as a reference geometry. 

For the metrics before  rotation (\ref{vielbeinbefore}--\ref{f3old}) we have $N^\mu_t=0$,  $N_t= \sqrt{|g_{00}|}= e^{\Phi/4} $,  $dS_t^\infty =\frac{1}{8} e^{2( \Phi + g + h ) + k} $, $\ n^\mu= \sqrt{g^{rr}}\delta^\mu_r=e^{-\Phi/4}e^{-k}$. The extrinsic curvature is
\begin{align}
^8 K &= \nabla_\mu n^\mu = \frac{1}{\sqrt{g_{_9}}}\partial_\mu(\sqrt{g_{_9}} n^\mu)= e^{-\Phi/4 -k} \Big[ 2(\Phi' + g' + h') + k'\Big], 
\end{align}
where $g_{_9}$ denotes the determinant of the 9-dimensional constant time slice $\Sigma_t$. The requirement that the induced metrics on $S_t{^\infty}$ agree at the boundary implies\footnote{The subscripts $ns$ and $su$ stand for non-supersymmetric and supersymetric respectively.}, 
\begin{align}\label{eq:match1-bdy}
	e^{\frac{\Phi_{ns}}{2}} e^{ 2 g_{ns}}&=e^{\frac{\Phi_{su}}{2}} 
e^{ 2 g_{su}},\quad 
e^{\frac{\Phi_{ns}}{2}} e^{ 2 h_{ns}}=e^{\frac{\Phi_{su}}{2}} e^{ 2 h_{su}},
\quad 
e^{\frac{\Phi_{ns}}{2}} e^{ 2 k_{ns}}=e^{\frac{\Phi_{su}}{2}} e^{ 2 k_{su}},
\quad 
\end{align}
and the $g_{00}$ component agrees if
\be\label{eq:match2-bdy}
e^{-\frac{\Phi_{ns}}{2}} = e^{-\frac{\Phi_{su}}{2}} .
\ee
All the quantities in \eqref{eq:match1-bdy} and \eqref{eq:match2-bdy} are evaluated at some large but finite $r_c$ that acts as a cutoff.
Using \eqref{eq:hawking-horowitz}, the energy is
\begin{align}\label{eq:energy}
	E&=-\frac{ vol_8}{64 \pi} \lim_{r_c \to\infty}\left\{e^{-k_{ns}}\left(e^{2\Phi_{ns} 
+2 g_{ns} + 2 h_{ns} + k_{ns}}\right)' - e^{-k_{s}}\left(e^{2\Phi_{s} + 2 g_{s} + 
2 h_{s} + k_{s}}\right)'  \right\}.
\end{align}	
Before evaluating \eqref{eq:energy} we have to satisfy the matching conditions 
at the boundary, \eqref{eq:match1-bdy} and \eqref{eq:match2-bdy}. 
In order to do this we have to use the 
most general asymptotics of a   supersymmetric solution.
As discussed in eq.(\ref{recoverSUSYUV}), analyzing the BPS equations (see Appendix A in 
reference \cite{Casero:2006pt}) we see that the most 
general supersymmetric UV asymptotics is obtained by replacing
\be\label{eq:susyrules}
W_{20} \ra 0,\ V_{40} \ra
2 e^{2 \rho_o} (1 + Q_o),\ H_{11}\ra 1/2,\ \Phi_{30} \ra 
-3\frac{(3 + 4 Q_o)}{4 c_{+}^2 }e^{4 \Phi_\infty}  
\ee
in the  non-supersymmetric expansion \eqref{UVexpansionnonsusy}. 
Notice that this substitution  restores  the integration constants 
$Q_o, \rho_0 $ and $e^{\Phi_\infty}$ that are usually  
set to $-1,\  0$ and $1 $ respectively \cite{HoyosBadajoz:2008fw}. 
Reintroducing the integration constants is equivalent to using  
the shift invariance of the $r$ coordinate 
(encoded in $Q_0 $ and $\rho_0$ ) and the dilaton \cite{Gubser:2001eg}. 
Adjusting these  constants will allow us to satisfy 
the matching conditions at the boundary and cancel divergences in the energy. 
Given the complexity of the UV expansions 
the matching procedure is cumbersome but straightforward. 
Working to linear order in $W_{20}$ we obtain,
\be\label{eq:energy-before}
E=\frac{1}{24 \pi}c^2_{+} e^{2\rho_0 + 2\Phi_\infty} W_{20}.
\ee

After the  duality transformations the  UV asymptotics changes drastically. 
In this case we have $N_t^\mu=0$, $\ N_t=\sqrt{|g_{00}|}
=\frac{ e^{-\Phi/4}}{H^{1/4}}$, $\ dS_t^\infty = \frac{1}{8} e^{ 3\Phi + 2g + 2h + k} H^{1/2}$, $\ n^\mu=  e^{-3\Phi/4} e^{-k}H^{-1/4}$ and
\be 
^8 K =  \nabla_\mu n^\mu= \frac{1}{\sqrt{g_{_9}}}\partial_\mu(\sqrt{g_{_9}} n^\mu)
=\frac{ e^{-\frac{ 3\Phi}{4} - k }}{2 H^{5/4}}\left[H' + 2 H( 3\Phi' + 2g' +2 h' + k')\right].
\ee
Note that here we  defined  $H = e^{-2\Phi} -  e^{-2 \Phi_\infty}$.
The regularized  energy after the rotation is 
\be\label{eq:energy-after-gen}
E=- \frac{vol_8}{64 \pi^2} \lim_{r\to\infty}\left\{\Delta_{ns} -\Delta_{su} \right\}
\ee
where
\be\label{eq:energy-after}
\Delta\equiv \frac{e^{-\Phi -k}}{\sqrt{H}}\left(\sqrt{H} e^{3\Phi + 2g + 2h + k}\right)'.
\ee
The matching conditions now read,
\begin{align}\label{eq:match1-after}
	& H_{ns}^{1/2}e^{\frac{3\Phi_{ns}}{2}+ 2 g_{ns}} = H_{su}^{1/2}e^{\frac{3\Phi_{su}}{2} + 2 g_{su}},\quad\quad H_{ns}^{1/2}e^{\frac{3\Phi_{ns}}{2} + 2 h_{ns}}=H_{su}^{1/2}e^{\frac{3\Phi_{su}}{2}+ 2 h_{su} },\quad\quad\nn\\
	& H_{ns}^{1/2} e^{\frac{3\Phi_{ns}}{2}+ 2 k_{ns}}= H_{su}e^{\frac{3\Phi_{su}}{2}+ 2 k_{su}},\quad\quad H_{ns}^{-\frac{1}{2}}e^{-\frac{\Phi_{ns}}{2}} = H_{su} ^{-1/2}e^{-\frac{\Phi_{su}}{2}}.
\end{align}
Note that 
\begin{align}\label{eq:beforeplusextra}
	\Delta&=e^{-k} \left(e^{2\Phi + 2g + 2h + k}\right)'  
+\frac{ e^{\Phi +2 g+ 2h}}{\sqrt{H}} \left(e^\Phi \sqrt{H}\right)'\nn\\
	&=\Delta^{before} + \Delta^{extra}
\end{align}
where $\Delta^{before}\equiv e^{-k} \left(e^{2\Phi + 2g + 2h + k}\right)'$ 
and $\Delta^{extra}\equiv \frac{ e^{\Phi +2 g+ 2h}}{\sqrt{H}} \left(e^\Phi \sqrt{H}\right)'$.
We have
\be\label{eq:energy-after-delta}
E=- \frac{vol_8}{64 \pi^2} \lim_{r_c\to\infty}
\left\{(\Delta^{before}_{ns}-\Delta^{before}_{s}) 
-(\Delta^{extra}_{ns}-\Delta^{extra}_s) \right\},
\ee
where all the functions are evaluated at some large but finite 
cutoff $r_c$. After adjusting the parameters to ensure that the 
induced metrics at the boundaries are the same, as required in \eqref{eq:match1-after}, 
we take the cutoff to  infinity. 
The first two terms in \eqref{eq:energy-after-delta} are the same as in the energy before rotation \eqref{eq:energy}. We find that --- to first order in $W_{20}$ --- the matching conditions are satisfied  using  the same set of integration constants as before the rotation. Thus, the first two terms in \eqref{eq:energy-after-delta} give exactly the energy before rotation. Any difference in energies will come from the $extra$ terms  \eqref{eq:energy-after-delta}. However, it can be shown that using the integration constants necessary to satisfy \eqref{eq:match1-after},
\be
\lim_{r_c\to\infty}\left\{(\Delta^{extra}_{ns}-\Delta^{extra}_s) \right\}=0.
\ee
Thus the energy 
before and after rotation are the same\footnote{This suggests that 
the ADM Energy is `uncharged' under the U-duality, like probably are
also uncharged various thermodynamical quantities.}. 
Indeed, plugging in the UV expansions 
directly in \eqref{eq:energy-after-gen}  we  obtain, 
\be
E=\frac{1}{24 \pi}c^2_{+} e^{2\rho_0 + 2\Phi_\infty} W_{20}.
\ee
A couple of comments are in order. First, note that the overall constant that appears in the energy can be changed by shifting the value of the dilaton at infinity. Thus, the physically meaningful statement is that the energies before and after rotation have the same 
functional dependence on the parameters,  
\beq
E_{before}\sim E_{after} \sim 
c^2_{+} e^{2\rho_0 + 2\Phi_\infty} W_{20}.
\label{energyafterbeforezzz}
\eeq
Second, this calculation can be carried out to higher order in the SUSY breaking parameter  $W_{20}$. The divergences in the energy can be cancelled by subtracting an appropiate SUSY background. However, at higher orders there will always be a  discrepancy  of order $W_{20}^2$  of  the metrics at the 
boundary. This clearly indicates that 
the treatment presented in this section is valid only for soft supersymmetry breaking with small breaking parameter, $W_{20}$. Had we not expanded  around $W_{20}\sim 0$ the mismatch at the 
boundary could be arbitrarily large indicating that the non-supersymmetric solution does not approach the SUSY solution fast enough for the energy to be finite. Note that this substantiates the smallness of  $W_{20}$  seen numerically in the previous section  where the solutions  found have $W_{20}\sim \mathcal{O}( 10^{-5})$.

\section{Field Theory Aspects}\label{seccionqft}
\setcounter{equation}{0}
In this section we will analyze various field theory aspects 
of a non-SUSY version of the quiver that we called 
field theory I and described below eq.(\ref{globalsymm})
To this end, we will use the non-SUSY background one obtains
when plugging our numerical solutions in Section \ref{seccionnumerica}
in the background of eq.(\ref{configurationfinal}) 
dual to the field theory I.

To begin with, notice that in  eq.(\ref{configurationfinal}) we did not specify
the NS potential $B_2$. Since this will be useful below, we discuss it here 
(the result is different from the SUSY one). 

Following the intuition gained in the SUSY example,
we propose a $B_2$ of the  form
\begin{align}
	B_2=	b_1(\rho) e^{\r 3} + b_2(\rho) e^{\theta\varphi} + b_3(\rho) e^{12} + b_4(\rho) e^{\theta2} + b_5(\rho) e^{\varphi 1},
\label{b2zzz}
\end{align}
by imposing that $dB_2=H_3$ and that the Page charge vanishes
$Q_{\text{Page, D3}}=0$ 
(see below) we obtain --- all details 
are discussed in Appendix \ref{B2appendix} --- 

\begin{align}
	b_1	&=		\frac{e^{2g-2k}}{4\hat{h}} \left[		2b_3 \Phi' - 3\hat{h}b_3 \Phi'  - 4\hat{h}b_3 g' - 2\hat{h}b_3'
																								+\k N_c e^{\frac{3\Phi}{2}-2h} \hat{h}^{\frac12} \left(	a^2 - 2ab +1	\right)
																				\right]	\nn\\
	b_2 &=		\frac{e^{-2h}}{4\hat{h}^{1/2}} \left\{	e^{2g}\hat{h}^{\frac12} \left(  1-a^2  \right)b_3
																											-\frac{\k}{N_c} e^{\frac{3\Phi}{2}} \left[	N_c^2 (a-b) b' + 4e^{2(g+h)} \Phi'
																																							\right] 
																							\right\} \nn\\
	b_4 &= b_5
			=			-\frac12 e^{g-h} a b_3		-		\frac{	\k N_c e^{\frac{3\Phi}{2}-g-h} b'	 }
																							 {4\hat{h}^{1/2}},  
\label{eq:bs}
\end{align}
with $b_3(\r)$ an undetermined function. 
This freedom corresponds to a gauge transformation.
A general $B_2$ can be expressed as
\begin{align}
B_2 = (B_2)_{b_3 = 0} -\frac{1}{2} d\!\left(  e^{2g-k+\Phi/4} \hat{h}^{1/4} b_3 \  e^3     \right).
\end{align}
Before computing various observables of 
the strongly coupled non-SUSY field theory I, we will quote another
quantity that will appear frequently in
the analysis.
This is a periodic 
quantity in the string theory. Given
the two cycle defined as,
\beq
\Sigma_2=[\theta=\tilde{\theta}, \varphi=2\pi-\tilde{\varphi}, \psi=\psi_0],
\eeq
we define
\beq
b_0=\frac{1}{4\pi^2}\int_{\Sigma_2}B_2.		\label{b0def}
\eeq
When computed explicitly using the form of the $B_2$ potential
in eqs.(\ref{b2zzz}-\ref{eq:bs}), we obtain
\beq
b_0(\psi_0)= \frac{\k N_c}{4\pi}e^{2\Phi}b'(b+\cos\psi_0) 
-\frac{\k}{\pi N_c} e^{2\Phi+2h+2g}\Phi'
\label{b0zz}
\eeq
These quantities together with those appearing in the background 
of eq.(\ref{configurationfinal}) will be important in the study of 
the non-perturbative
field theory dynamics.
\subsection{Calculation of observables}
We now move into the calculation 
of observables that will help us understand the field theory
interpretation of our solution.
\subsubsection{Interesting Asymptotic Behaviors}

We start by studying some combination of fields that
have a particular behavior. For this it is convenient to 
reduce the system to five dimensions
as was done in \cite{Elander:2011mh}.
Once in five dimensions,  the paper 
\cite{Elander:2011mh}
shows that 
some of the 5-d fields are invariants under the rotation. 
These fields are the dilaton $\Phi$ and the combinations
\beq
M_1=1+a^2+ 4 e^{2h-2g},\;\;\;M_2=  e^{2h+2g-4k}
\eeq
The corresponding UV expansions are 
(see Section \ref{seccionnumerica} and 
Appendix \ref{explicitexpansionsappendix} for the notation)
\begin{align}
	M_1   &=   2+ \left(8 H_{11} \rho +3 c_+ W_{20}^2+2 Q_o\right)
\frac{e^{-4 \rho /3}}{c_+} + 
O\bigl(e^{-8 \rho /3}\bigr),    \nn \\ 
	M_2   &=   \frac{9}{16}-\frac{27}{16} W_{20}^2 e^{-4 \rho /3}+ 
+O\bigl(e^{-8 \rho /3}\bigr) .
\end{align}

%
%
Now, suppose that we define a variable $z=e^{-2\r/3}$.
Any field ${\cal M}$ that for $z\to 0$ scales 
like ${\cal M}\sim z^\Delta$ 
indicates either the insertion of a relevant/marginal operator
or the VEV for an operator of dimension $\Delta$ (if $\Delta>0$ 
or $\Delta=0$). On the other hand, if $\Delta<0$,
it indicates the insertion in the Lagrangian 
of an irrelevant operator of dimension $(4-\Delta)$.

Using the UV expansion of the dilaton 
(see Appendix \ref{explicitexpansionsappendix}), 
we see that the dilaton corresponds to a 
marginal operator of dimension $\Delta=4$
(this is identified with a combination of 
gauge couplings $g_+^2$ discussed below).
The expansion of the function $b(\r)$ --- see again 
Appendix \ref{explicitexpansionsappendix} ---
indicates that the SUSY
breaking constant $W_{20}$ corresponds to the insertion of an operator 
of dimension three in the lagrangian. We associate this operator with
the mass for the gaugino and in an analogous way, 
the constant $e^{2\r_o}$ is 
associated with the VEV for the gaugino. The association is not exact, in the sense 
that once SUSY is broken there could be a contribution of $W_{20}$ to the gaugino VEV.
Then, schematically we have
\beq
W_{20}\to m \lambda\lambda,\;\;\; e^{2\r_0}\to \langle\lambda\lambda\rangle\sim \Lambda_{YM}^3.
\label{mapxxxx}
\eeq
Following this logic, the expansion of the field 
$M_1\sim z^2$ is interpreted as the VEV for 
a dimension two operator \cite{Dymarsky:2005xt},
\beq
{\cal U}\sim tr[AA^\dagger - B^\dagger B].
\eeq
This same operator gets a VEV in the SUSY case and is the one that
allows us to explore  the baryonic branch. Notice that 
the SUSY breaking coefficient $W_{20}$ contributes
to this VEV.

In the theory of type I, that has two gauge groups, we should expect two independent
gaugino masses. Here, the solution is obtained by U-duality applied on
a background dual to a Theory of type II, with only one gauge group.
We seem to have only one integration constant associated with gaugino mass, that is $W_{20}$.
As emphasized below eq.(\ref{constzzzz}) 
the numbers $V_{21}, W_{21}$ corresponding to
behavior of the functions $a\sim b\sim \r e^{-2\r/3}$, which could be associated
with this second mass parameter, turn out to vanish in our particular solution.

\subsubsection{Energy}
We take the expressions for the ADM Energy of the non-SUSY
backgrounds as derived in eqs.(\ref{eq:energy-before}) and (\ref{energyafterbeforezzz}) 
and we
use the map described in eq.(\ref{mapxxxx}), we obtain that
\beq
E_{ADM}\sim c_+^2 e^{2\Phi(\infty)} e^{2\rho_0}W_{20}\sim m \Lambda_{YM}^3.
\eeq
Then the energy is proportional to the gaugino mass 
and the strong coupling scale, as expected.
\subsubsection{Charges}\label{mpcharges}
We will define
the Maxwell and Page Charges
\bea
& & Q_{\text{Maxwell, D3}}=\frac{1}{16\pi^4}\int_{X_5} F_5,\;\;\;\;\;
Q_{\text{Maxwell, D5}}=\frac{1}{4\pi^2}\int_{X_3} F_3,\nonumber\\
& & Q_{\text{Page, D3}}=\frac{1}{16\pi^4}\int_{X_5} F_5- B_2\wedge F_3, \label{eq:chargedefs}
\eea
where the manifold $X_5=[\theta,\varphi,\tilde{\theta},\tilde{\varphi},\psi]$
and $X_3=[\tilde{\theta},\tilde{\varphi},\psi]$.
As in the SUSY case we have that
\begin{align}
	Q_{\text{Maxwell, D3}} = \frac{\k}{\pi} e^{2g+2h+2\Phi}\Phi' ,		\qquad\qquad
	Q_{\text{Maxwell, D5}} = N_c. 
\end{align}
We have also imposed that $Q_{\text{Page, D3}}=0$ in determining the $B_2$
field of eq.(\ref{b2zzz}) --- see Appendix \ref{B2appendix} for details. 
The vanishing of the D3-Page charge is a feature of
the SUSY non-singular solutions; this is the reason why we imposed it here.
It would be interesting to see if one can obtain
a regular non-SUSY  solution in
the presence of sources indicated by a non-vanishing  
Page charge.
Using the UV expansions, the Maxwell charge for D3 branes
is
\begin{align}
 Q_{\text{Maxwell, D3}} 
&= \frac{e^{\Phi_\infty}}{\pi}\r -
\frac{1}{24\pi}\left(9 e^{\Phi_\infty }
+ 4 c_+^2 e^{-3\Phi_\infty} \Phi_{30}   \right)\nonumber\\ 
&  \qquad + \frac{33e^{\Phi_\infty}W_{20}^2}{32\pi} e^{-4\rho/3}
+O\bigl( e^{-8\rho/3} \bigr).		
\end{align}
So, we see that $W_{20}$,
the same number that determines the mass of the gaugino
according the discussion above, changes the large energy
value of the Maxwell charge
(correspondingly of the c-function --- see below) in a  
subleading way, as expected.

\subsubsection{Gauge couplings and beta functions}
Let us review briefly what happens in the SUSY case.
In the $SU(N_c+n)\times SU(n)$ SUSY quiver, we have two couplings
$g_1, g_2$. Close to the Klebanov-Witten conformal point 
(in the UV), the anomalous dimensions
are $\gamma_{A,B}\sim -\frac{1}{2}$. 
This implies that the beta functions for the diagonal combinations
\bea 
\beta_{\frac{8\pi^2}{g_{-}^2}}
=
\beta_{\frac{8\pi^2}{g_{1}^2}}-
\beta_{\frac{8\pi^2}{g_{2}^2}}= 6 N_c,\;\;\;
\beta_{\frac{8\pi^2}{g_{+}^2}}
=
\beta_{\frac{8\pi^2}{g_{1}^2}}+
\beta_{\frac{8\pi^2}{g_{2}^2}}=0.
\eea
As in the SUSY case, we will adopt the 
definitions\footnote{These are 
strictly correct in the $N=2$ examples and the KW fixed point. 
We adopt the definition here to get a handle on the non-SUSY dynamics.}
\beq
\frac{4\pi^2}{g_+^2} = \pi e^{-\Phi},		\qquad\qquad
\frac{4\pi^2}{g_-^2} = 2\pi e^{-\Phi} \left[ 1 - b_0(\pi) \right]
\eeq
where $b_0(\psi_0)$ is defined in eq. (\ref{b0def})-(\ref{b0zz}).
We obtain 
\begin{align}
	\frac{4\pi^2}{g_-^2} &=  2e^{-\Phi} \left( \pi  + \frac{\k}{N_c} 
e^{2g+2h+2\Phi}\Phi' \right) - \frac{\k N_c}{2} e^{\Phi} (b-1) b',
\end{align}
Notice that the result is 
independent of the gauge artifact function $b_3(\r)$.
In the UV, these formulas are typically trustable. 
The explicit expansions are
\begin{align}
\frac{4 \pi ^2}{g_+^2}  
=  e^{-\Phi_\infty}\pi+\left(\frac{3e^{-\Phi_\infty}\pi}{2c_{+}^{2}}
\rho-\frac{1}{4}e^{-5\Phi_\infty}\pi\Phi_{30}\right)e^{-8\rho/3}
+O\bigl(W_{20}^{2} e^{-4 \rho }\bigr)
\end{align}
and
\begin{align}
\frac{4 \pi ^2}{g_-^2}  
				&=   \left(2 \rho-\frac{1}{3} c_+^2 
\Phi _{30} e^{-4 \Phi_\infty} +2 \pi  
e^{-\Phi_\infty}-\frac{3}{4}\right)
						-\frac{3}{4} W_{20} e^{-2 \rho /3}
						+O\bigl(e^{-4 \rho /3}\bigr). 
\end{align}
Let us now compute the beta functions as read from the geometry.
We will use the radius/energy relation
\beq
r= e^{2\r/3}=\frac{\mu}{\Lambda}
\eeq
where $\mu$ is the energy scale at which we probe the process and $\Lambda$
the reference or strong coupling scale of the given gauge group.
Notice that this choice is arbitrary, just reflecting the possibility 
of choosing a scheme. Other monotonic relations $\r(\mu)$ would  
express the beta function
in other schemes.
To calculate the beta functions we perform
\bea
& & \beta_{\frac{8\pi^2}{g_{-}^2}}=
\frac{d}{d\r}\left(\frac{8\pi^2}{g_{-}^2}\right)\frac{d\r}{d\log(\mu/\Lambda)}=
6 N_c + W_{20}N_c \frac{\Lambda}{\mu},\nn\\
& & \beta_{\frac{8\pi^2}{g_{+}^2}}= 
\frac{d}{d\r}\left(\frac{8\pi^2}{g_{+}^2}\right)\frac{d\r}{d\log(\mu/\Lambda)}=
O\left(\log\left(\frac{\Lambda}{\mu}\right) \frac{\Lambda^4}{\mu^4}\right) .  
\label{betas}
\eea
We have reinstated the factor of $N_c$ in the expansions.
With a naive use of the NSVZ expression for the Wilsonian
beta functions one may have interpreted this result for $\beta_-$ 
as the SUSY breaking parameter 
$W_{20} $  changing 
slightly the value of the anomalous dimensions
$
\gamma_{A,B}\sim -\frac{1}{2}+ O\left(W_{20}\frac{\Lambda}{\mu}\right).
$
But this is not matching with the analogous calculation for $\beta_+$.
Hence this solution does not respect the NSVZ expression (as expected).
Also, notice that while in the SUSY case, 
the beta functions receive corrections
$O\left(\frac{\Lambda^3}{\mu^3}\right)$, we have here an example 
where the SUSY breaking parameters produce lower order corrections 
$O\left(\frac{\Lambda}{\mu}\right)$.
Let us move now to IR observables.

\subsubsection{K-Strings}
We will follow the treatment in
\cite{Conde:2011rg}.
We need to evaluate the action for a D3 brane that extends on
the manifold
$\Sigma=[t,x_1, \theta=\tilde{\theta},\varphi=2\pi-\tilde{\varphi}].$
The D3 brane is sitting at $\r=0$ 
but can move on the angle $\psi$, 
so that it will minimize its energy. 
The (string frame) metric seen by such a D3 brane is
\begin{align}
ds_{ind}^2	&= \frac{e^{\phi_0}}{\sqrt{\hat{h}_0}}
							 \left\{ dx_{1,1}^2 + N_c \hat{h}_0 \frac{h_1}{2}\left[	 d\chi^2
																																		+	\sin^2\chi
																																			\left( d\theta^2 + \sin^2\theta d\varphi^2 \right)
																															\right]
 																	\right\}
\end{align}
where we have written $\psi = 2\chi+\pi$, and 
used the values of the functions at $\rho=0$:
\begin{align}
	e^{2g(0)}=e^{2k(0)}=\frac{h_1}{2}, \qquad e^{2h(0)}=0, \qquad \Phi(0)=\phi_0, \qquad a(0)=1.
\end{align}
We have additionally written $\hat{h}_0\equiv \hat{h}(0) = 1-e^{2\phi_0-2\Phi_\infty}$.

The RR field and its potential are,
\bea
& &F_3|_\Sigma = 2N_c \sin^2\chi \;\Omega_2 \wedge d\chi,\;\;
C_2|_\Sigma = N_c \left( \chi - \frac{\sin2\chi}{2} \right)
\Omega_2,\nonumber\\
& & \Omega_2=\sin\theta\ d\theta \wedge d\varphi.
\eea
Using eqs.(\ref{b2zzz}--\ref{eq:bs}) and the fact that $b'(0)=\Phi'(0)=0$ we find that the NS potential $B_2$ vanishes.

We will  turn on an electric field $F_2 = F_{tx}\ dt\wedge dx$ 
in the space-time directions.
Then the Born-Infeld-Wess-Zumino 
action gives an effective one dimensional lagrangian,
\bea
L_\text{eff}= - 4\pi T_{D3}N_c\left[   \frac{h_1}{2}	\sqrt{	e^{2\phi_0}-\hat{h}_0\frac{F_{tx}^2}{4}} \ \sin^2\chi- 
\left(\chi-\frac{\sin2\chi}{2}\right)F_{tx} \right].
\eea
This is equivalent to eq. (9.8) of \cite{Conde:2011rg}, 
with modifications which result from	
the U-duality,
\begin{align}
	\beta = \frac{h_1}{2} \to \frac{h_1}{2} \sqrt{\hat{h}_0},		\qquad\qquad
	e^{2\Phi(0)} \to \frac{e^{2\phi_0}}{\hat{h}_0}.    \label{kstringrotmods}
\end{align}
The rest of the discussion then proceeds as in \cite{Conde:2011rg}. 
We impose the equation of motion for $F_{tx}$ and quantize 
it to be an integer  multiple of the tension of the fundamental string,
$
\frac{\partial L_\text{eff}}{\partial F_{tx}}= \frac{k T_f}{2}$.
The resulting tension follows an {\it approximate}
sine-law, as in the whole baryonic branch, 
including the KS solution. This also happens for D5 solutions
in section 8 of reference \cite{Casero:2006pt}.

The influence from the SUSY breaking parameters enters 
only through the modifications \eqref{kstringrotmods}.
\subsubsection{The Non-SUSY Seiberg-like duality.}
\label{seibergsection}
We will follow the treatment in the SUSY case, as developed 
in \cite{Benini:2007gx}. 
The basic idea is go back to the quantity $b_0(\psi)$,
computed as specified around eq.(\ref{b0zz}) and compare
with what occurs in the SUSY case.
The Seiberg duality is identified with a large gauge transformation
such that $b_0\to b_0\pm 1$ and the charge of D3 branes changes by 
$\pm N_c$.

Consider  the Page charge od Section \ref{mpcharges}; 
a large gauge transformation
on $B_2$ will change $b_0$ in one unit. This translates in the change of $N_c$
units in the Page charge. This works exactly as in \cite{Benini:2007gx}.

Let us now study how the Maxwell charge `sees' the Seiberg duality.
We will focus on the UV part of the background, where the cascade is known to
work in the SUSY case.
Following the steps  described in  
Appendix \ref{seibergappendix}, 
we have
\beq
b_0=\frac{\hat{h}^{1/2} e^{\Phi/2}}{\pi}
\Big[ b_2 e^{2h}- b_4(a+\cos\psi_0) e^{h+g}   \Big]=
\frac{N_c}{\pi}[(f+\tilde{k}) + (\tilde{k}-f)\cos\psi_0]
\eeq
with (using the explicit values for $b_2, b_4$)
\bea
& & f= \frac{e^{\Phi/2}\hat{h}^{1/2}}{2N_c}[b_2 e^{2h}- 
b_4 e^{g+h}(a-1)]=\kappa\frac{e^{2\Phi}}{8}[b'(b-1) -
\frac{4}{N_c^2}e^{2g+2h}\Phi'],\nonumber\\
& & \tilde{k}= \frac{e^{\Phi/2}\hat{h}^{1/2}}{2N_c}[b_2 e^{2h}- 
b_4 e^{g+h}(a+1)]= \kappa\frac{e^{2\Phi}}{8}[b'(b+1) -
\frac{4}{N_c^2}e^{2g+2h}\Phi'], \nonumber\\
& &  \to b_0=  \frac{\k N_c e^{2\Phi}}{4\pi}b'(b+\cos\psi_0)	
-\frac{\k e^{2\Phi+2h+2g}}{\pi N_c}\Phi'       .
\eea
Now, it is interesting to notice that --- far in the UV --- the Maxwell charge 
\beq
Q_{Max, D3}=\frac{\k}{\pi}
e^{2g+2h+2\Phi}\Phi'=  
\frac{\k N_c^2 e^{2\Phi}}{4\pi}b'(b+\cos\psi_0) -  N_c b_0
\eeq
changes under a change in $b_0$ as,
\beq
b_0\sim b_0\pm 1 \to Q_{Max, D3}\sim Q_{Max, D3}\mp N_c.
\eeq
Specially, notice that for large values of $\r$
the `correction-term'  $b'(b+\cos\psi_0)$ is quite suppressed.
This `correction' 
is more suppressed in the SUSY case, where $b'\sim e^{-2\r}$,
in contrast to our non-SUSY solutions, where $b'\sim e^{-2\r/3}$. 
The `Seiberg duality', associated with a large gauge transformation of
index $k$ that changes the Maxwell charge in $kN_c$ units is  better
approximated in the SUSY than in the non-SUSY case.
Nevertheless, in both cases, the transformation is good at leading order.

So, as expected, far in the UV we could think that the decrease in the
Maxwell charge is interpreted as a non-SUSY version of 
Seiberg duality that is at work here.
\subsubsection{Domain Walls }
Let us compute the tension of a domain wall as the effective tension of
a five brane that sits at $\r=0$ and is extended along
$\Sigma_6=[t,x_1,x_2,\tilde{\theta},\tilde{\varphi},\psi]$.
Before the U-duality for field 
theories of type II, we use the background in eq.(\ref{f3old}) and obtain
that the 
induced metric on such five brane is (in string frame)
\beq
ds_{ind}^2= e^{\Phi}\Big[ dx_{1,2}^2+ \frac{e^{2g}}{4}(d\tilde{\theta}^2
+\sin^2\tilde{\theta}d\tilde{\varphi}^2)+ \frac{e^{2k}}{4}(d\psi+ 
\cos\tilde{\theta} d\tilde{\varphi})^2\Big]
\eeq
The induced tension on the three dimensional wall is
\beq
T_{eff}= \left. 2\pi^2 T_{D5} e^{2\Phi+ 2g+k}\right|_{\rho=0}
= \frac{\pi^2 T_{D5} e^{2\phi_0} h_1^{3/2}}{\sqrt{2}},
\label{tensiond5nonsusy}\eeq
which is unchanged from the SUSY result.

After the U-duality, in the background of eq.(\ref{configurationfinal}), 
we place 
a similar five brane, the induced metric is,
\beq
ds_{ind}^2= 
e^{\Phi}
\Big[ \frac{1}{\sqrt{\hat{h}}}
dx_{1,2}^2+\sqrt{\hat{h}}\Big(
 \frac{e^{2g}}{4}(d\tilde{\theta}^2
+\sin^2\tilde{\theta}d\tilde{\varphi}^2)+ \frac{e^{2k}}{4}(d\psi+ 
\cos\tilde{\theta} d\tilde{\varphi})^2\Big)\Big].
\eeq
There is also an induced $B_2$ field,
\beq
B_2= \frac14 \sqrt{\hat{h}} 
e^{2g + \Phi/2} b_3(\r)  
\sin\tilde{\theta} d\tilde{\theta}\wedge d\tilde{\varphi}.
\eeq
In order to have a gauge invariant Born-Infeld 
Action, we must add the $F_2$ field on
the world-volume of the brane. Indeed, the change due to a gauge
transformation of the $B_2$ field is cancelled by a 
(non-gauge)-transformation on $F_2$\footnote{One can also add a field strength
$F_2$ such that aside from cancelling the 
gauge-variance of $B_2$ adds a kind 
of `magnetic charge' to the domain wall or a 
Maxwell-like term in the Minkowski directions. 
We will not consider the addition of these extra components 
of $F_2$ as they will typically raise the energy of the wall.},
\beq
B_2\to B_2 + d\L_1,\;\;\; F_2\to F_2 - d\L_1.
\eeq
Hence, we need to turn on the worldvolume of the brane a
gauge field strength,
\beq
F_{\tilde{\theta}\tilde{\varphi}}=- \frac{\sqrt{\hat{h}}}{4}e^{2g+\Phi/2}
b_3(\r)\sin\tilde{\theta}.
\eeq
This implies that the BIWZ action will be
\beq
S = -T_{D5}(4\pi)^2 \frac{e^{2g+k+2\Phi}}{{8}}\int d^{2+1}x
\eeq
which gives the same effective tension  as in eq.(\ref{tensiond5nonsusy}) 
and the same as in the SUSY case \footnote{
Notice, that in the SUSY case we have (using eq. \ref{eq:SUSYB2})
\begin{align}
	b_3 = - \k e^{3\Phi/2}\hat{h}^{-1/2} \cos \alpha,
\end{align}
which vanishes for $\r=0$.}. 
Then the tension before 
and after the U-duality is the same. 
As a side remark, 
one may wonder if it is possible  to fix the value
of $b_3$ at $\r=0$ using some physical criterium. 
Though it is not an invariant quantity, the small $\r$ expansion of  
\beq
B_{\mu\nu}B^{\mu\nu}\sim \frac{b_3(0)^2}{\r^2} + \cdots
\eeq
suggests that we should take $b_3(0)=0$ as in the SUSY case.
\subsubsection{Central charge}
We will calculate the central charge of this non-susy solution. 
We should follow the usual procedure of
\cite{CentralCharge}, that requires a 
reduction to five dimensions. However, an equivalent treatment 
presented in
\cite{Klebanov:2007ws}
indicates that for any string-frame metric of the form
\beq
ds^2\sim \a(\r) dx_{1,d}^2 +\a(\r)\beta(\r) d\r^2+ g_{ij}(\r,y)dy^i dy^j
\eeq
we define
\beq
V_{int}=\int d\vec{y} \sqrt{\det g_{ij}},\;\;\;H=e^{-4\Phi}\a^d V_{int}^2.
\eeq
and the central charge (for $d=3$)
is given by
$c\sim \frac{\beta^{3/2} H^{7/2}}{(H')^3}
$, in our case
\beq
c\sim \frac{e^{2h+2g+2\Phi+4k} \hat{h}^2}{(2h'+2g'+2\Phi'+k' 
+\frac{\hat{h}'}{2\hat{h}})^3}
\eeq
In the IR, the explicit expansion for the central charge is
\begin{align}
c
&\sim e^{2\phi_{0}-4\Phi_\infty}\left(e^{2\Phi_\infty}-e^{2\phi_{0}}\right)^{2}h_{1}^{4}\rho^{5} 	\nn\\
&\qquad			+\frac{1}{9}e^{2\phi_{0}-4\Phi_\infty}\left(e^{2\Phi_\infty}-e^{2\phi_{0}}\right)h_{1}^{2}\bigg[e^{2\Phi_\infty}\left(-16-15h_{1}k_{2}+12h_{1}^{2}\right)		\nn\\
&\qquad\quad		+e^{2\phi_{0}}\left(28+15h_{1}k_{2}-12h_{1}^{2}+9v_{2}^{2}\right)\bigg]\rho^{7}+O\bigl(\rho^{9}\bigr), 
\end{align}
and in the UV we have
\begin{align}
c
&\sim e^{2\Phi_\infty}\rho^{2}-\left(\frac{3}{4}e^{2\Phi_\infty}+\frac{1}{3}c_{+}^{2}e^{-2\Phi_\infty}\Phi_{30}\right)\rho+O\biggl(\frac{1}{\rho}\biggr).
\end{align}
It is immediately clear that the SUSY-breaking parameters have no effect at the leading order in the UV. However, in the IR the question is more subtle. Although none of the explicit SUSY-breaking parameters appear in the leading term there is an effect. This is because in the SUSY case there are only two independent parameters, so that fixing $h_1$ and $\phi_0$ is sufficient to determine $\Phi_\infty$. In the non-SUSY case the discussion of sections \ref{parametercounting} and \ref{seccionnumerica} suggests that there is one more parameter, which breaks SUSY. This means that even with fixed $h_1$ and $\phi_0$ we can 
expect that $\Phi_\infty$ varies as a function of the 
SUSY-breaking parameter. Indeed, when in appendix \ref{NumApp} 
we compare SUSY and non-SUSY numerical 
solutions with the same  $h_1$ and $\phi_0$, we find that $\Phi_\infty$ changes.


\subsubsection{Force on a probe $D3$-brane}
We will now consider a D3 probe brane that extends in the 
Minkowski directions and is free to move in the radial
direction as suggested in \cite{Dymarsky:2005xt},
\beq
D3:[t,x_1,x_2, x_3],\;\;\; \r(t).
\eeq
the induced metric and RR four form field are obtained 
from the string frame version of eq.(\ref{configurationfinal}),
\bea
& & ds_{ind}^2=e^{\Phi}\hat{h}^{-1/2}\Big[-dt^2(1-\hat{h}e^{2k}\r'^2) + 
dx_{1}^2 + dx_2^2 + dx_3^2   \Big], \nonumber\\
& & C_4=-\k\frac{e^{2\Phi}}{\hat{h}}dt\wedge dx_1\wedge dx_2\wedge dx_3.
\eea
this gives an action for the D3 brane\footnote{Notice that in the 
way things have been defined, the action for a D3
has the WZ term with the same sign as the BI term. 
See eq.(2.13) in the paper \cite{GMNP}.},
\beq
S_{BIWZ}= -T_{D3}V_3\int dt \left(\frac{e^{\Phi}}{\hat{h}} 
\sqrt{1-\hat{h} e^{2k}\r'^2}-\k \frac{e^{2\Phi}}{\hat{h}}\right).
\eeq
We then approximate this for 
small velocities and change to the variable
$dr=e^{k+\Phi/2}d\r$ and get
\beq
S= T_{D3} V_3\int dt \left( \frac{r'^2}{2} - \frac{e^{\Phi}}{1+\k e^{\Phi}} \right).
\eeq
the force on this probe is then
\beq
f=\frac{e^{\Phi/2 -k}}{(1+\k e^{\Phi})^2}\Phi'.
\eeq
In the IR, the explicit expansion for this force is
\begin{align}
f=\frac{2\sqrt{2}}{3} \frac{e^{\frac{\phi_{0}}{2}+2\Phi_\infty}\left(4+3v_{2}^{2}\right)}{\left(e^{\phi_{0}}+e^{\Phi_\infty}\right)^{2}h_{1}^{5/2}}\rho+O\bigl(\rho^{2}\bigr), 	
\end{align}
and in the UV we have
\begin{align}
f=\left[\sqrt{\frac{3}{2}}\frac{e^{\frac{\Phi_\infty}{2}}}{c_{+}^{5/2}}\rho-\frac{e^{\frac{-7\Phi_\infty}{2}}\left(9e^{4\Phi_\infty}+4c_{+}^{2}\Phi_{30}\right)}{8\sqrt{6}c_{+}^{5/2}}\right]e^{-10\rho/3} +O\bigl(e^{-14\rho/3}\bigr).
\end{align}
As expected, the force vanishes quickly in the far UV, where the solution
approaches the KS background. 
Also, notice that in the radial coordinate $r\sim e^{-2\r/3}$, 
the force is $f\sim \frac{\log r}{r^5}$ as obtained in \cite{Dymarsky:2011pm}.
The SUSY breaking parameters do not influence this small force.
In the other hand, the breaking of SUSY  explicitly 
changes the value of the force in the IR, as expected.
\subsection{Field theory comments}
This section  relies on the ideas of 
\cite{Masiero:1984ss}-\cite{Aharony:1995zh}, but most 
fundamentally
 on the analysis of the paper
\cite{Aharony:1995zh}. Similar ideas that may be useful 
in thinking about our string backgrounds
have been put forward for example in
\cite{Evans:1995ia}. This paper studies  
non-SUSY deformations of ${\cal N}=1$ SQCD.
We use this to analyze the quiver field theory of type I. 
This is as we discussed, a non-SUSY deformation
of the KS-quiver. In the SUSY case, the KS-field theory
can be understood as N=1 SQCD with gauged flavor group and a 
quartic superpotential (see for example
\cite{Strassler:2005qs}) and due to this, the results of \cite{Aharony:1995zh}
are important to us.
The qualitative results of the paper \cite{Aharony:1995zh} 
become quantitatively
accurate once we take the SUSY breaking parameters much smaller than the
relevant scale of the problem, 
namely $\Lambda_{SQCD}$\footnote{Ofer Aharony explained us
that the soft breaking mass terms for squarks could have different signs
under a Seiberg Duality, see \cite{ArkaniHamed:1998wc}. 
This technical subtlety 
seems to play no role in our analysis}. In our case, 
this is reflected in the smallness of the coefficient $W_{20}$.

In this case, lots of the structure of Seiberg's SQCD 
\cite{Seiberg:1994pq}
remains. 
Particularly interesting
to us is the fact that for $SU(N_c)$ SQCD with $N_f$ flavors and with 
$N_f=N_c$, there exists a vacuum which breaks spontaneously
the $U(1)$-baryonic symmetry and this vaccum persists in 
the non-SUSY analysis of \cite{Aharony:1995zh}.
This will be relevant for us as the case $N_f=N_c$ is associated in 
the SUSY case with the last 
step of the cascade. We then argue that our geometry 
describes a situation where SUSY is broken by gaugino masses and other VEV's
and the baryonic symmetry
is broken by the vacuum state.

In a bit more detail, the authors of \cite{Aharony:1995zh}, 
added to the SQCD lagrangian a term of the form
\bea
& & L= L_{SQCD}+ \Delta L,\nonumber\\
& & \Delta L\sim \int d^4\theta M_Q(Q^\dagger e^{V}Q +\tilde{Q}^\dagger e^{-V}\tilde{Q}) 
+\int d^2\theta M_g S,
\eea
where $S$ is the superfield $S=Tr[W_\a W^\a]$, $M_Q$ is a vector multiplet whose D-component equals the mass of the squarks ($- m_q^2$)
and $M_g$ is a chiral multiplet whose F-component is the mass of the gluino.
The  authors of 
\cite{Aharony:1995zh} argued that to 
leading order in the SUSY breaking parameters $M_Q, M_g$ one can write an effective lagrangian in terms of 
mesons $\hat{M}$, baryons ($B, \tilde{B}$) and $S$,
\beq
\Delta L\sim \int d^4\theta B_M M_Q tr[\hat{M}^\dagger \hat{M}] + B_b M_Q 
(B^\dagger
B +\tilde{B}^\dagger \tilde{B})+\int d^2\theta M_g S+....
\label{nonsusydeformation}\eeq
The idea is then that one should supplement the usual actions 
and superpotentials discussed in the SUSY case with the SUSY breaking terms above.
In particular, in the case $N_f=N_c$ we will need to minimize
the potential term coming from eq.(\ref{nonsusydeformation}) together with
the potential coming from the SUSY superpotential
\beq
W= W_{tree}+W_{quant}=\kappa Tr (\hat{M}^\dagger \hat{M})+
\xi (\det M - \tilde{B}B -\Lambda^{2N_c}).
\eeq
Therefore, the vacua of the theory are those
that minimize the potential coming from the tree 
level superpotential, together with that from the SUSY breaking term, all subject to the constraint
in $W_{quant}$.
The result is that in the non-SUSY case, one finds one vacuum state where 
the baryons get a VEV and the mesons are at the origin of the moduli space,
$\hat{M}=0$.

In this way, we have argued that our solution, which breaks 
SUSY due to masses for the gauginos has  very similar behavior to
the KS-cascade (actually to the baryonic
branch in \cite{BGMPZ}).
We found  that many non-perturbative aspects behave very similarly
as the SUSY case: the expression 
of the domain wall tension is basically the same as in the SUSY case.
Of course, numbers will differ as the 
functions in the IR pick the influence of the
SUSY breaking terms. 
The tensions for k-strings gives an approximate 
sine-law, again with the SUSY breaking entering the value of the tension.
In the UV, the beta function for
the gauge couplings of the quiver and the leading order of the central charge
behave at leading order in the UV like
their SUSY counterpart, but in the  case of the beta functions, the
first correction is purely coming from SUSY breaking contributions.
The Seiberg duality (identified 
here with the change of the Maxwell charge 
under large gauge transformations of the NS B-field) 
behaves very approximately as in the SUSY
case. 
One can probably make an argument for self-similarity as 
presented in \cite{Strassler:2005qs}.

Obviously, what happens is that the SUSY breaking terms, for example 
the gaugino mass indicated by the quantity $W_{20}$, are 
not important at high energies. They enter some IR observables, 
correcting but not changing the 
qualitative
behavior expected from the SUSY example.
This suggests that we need
to think that our SUSY breaking scales are 
smaller than our strong coupling scale. Hence, 
the phenomena are the same as in the SUSY models,
but numerically there will be differences.
All this is in line with the analysis of 
\cite{Masiero:1984ss}-\cite{Aharony:1995zh}.

\section{Conclusions}
\setcounter{equation}{0}
In this paper we have used analytical (UV and IR series expansions) and numerical methods
to construct smooth backgrounds   dual to particular non-SUSY field theories.
The field theories in question can be thought of as softly-broken-SUSY versions of the 
field theories appearing in twisted D5 branes and
Klebanov-Strassler quivers.

We presented some details of the derivations, involving a  U-duality,
a careful numerical procedure and a detailed study of many  observables
at low and high energies. All this supports  the field theory interpretation
discussed towards the end of
Section \ref{seccionqft}. In other words,
the dynamics is basically the SUSY one, but with interesting details and 
deviations
coming from the soft-breaking terms.

Various things come to mind that would be nice to study using 
the backgrounds presented here.
Before the U-duality, it would be nice to study the effect 
on k-string tensions, domain walls
and the confining behavior of the Wilson loop, as there exist in the 
bibliography 
various results for  ${\cal N}=1$
Super-Yang-Mills being 
softly broken. Also, it would be nice to study the effect of
the mass terms responsible for the
breaking on the glueball spectrum.
It would also be interesting to see 
if our solutions can be of any help for the interesting
line of metastable broken SUSY, in the sense of providing a good set of UV
boundary conditions that break SUSY. This may be used 
to get ideas on the singular IR
behaviors obtained in \cite{Bena:2009xk}, \cite{Dymarsky:2011pm}.

It would be interesting to 
calculate the mass spectrum and compare it with
the result of the analogous glueballs in the KS/baryonic branch solution
\cite{Caceres:2000qe}. Also, it would be nice to find 
numerically
the expression
for the massless glueball corresponding to the spontaneous
breaking of the baryonic symmetry. The spectrum of mesons is also  
of interest.  In particular, comparing  the masses of the 
lightest scalar meson and the lightest  
vector meson we could learn  if the non-SUSY background presented 
here provides a plausible  holographic dual of nuclear 
forces \cite{Kaplunovsky:2010eh},\cite{Ihl:2010zg}.

Another problem that we are not addressing here: it is known that
taking the integration constant $c_+\to\infty$ leads, 
in the SUSY case, to the Klebanov-Strassler background (see \cite{GMNP}).
It is interesting to see this working numerically and to compare the solutions
found here --- in the limit --- with those found in the past
by first order fluctuations of the KS system (see \cite{Kuperstein:2003yt}).
In this line, a nice problem would be to find
the recent solution by Dymarsky and Kuperstein 
\cite{anatoly} as a scaling of ours.

\section{Acknowledgments:}
We would like to thank various colleagues for correspondence, 
nice discussions
and their valuable input to improve the presentation of this paper: 
Ofer Aharony, Adi Armoni, Stefano Cremonesi, Anatoly Dymarsky, 
Prem Kumar, Oscar Loaiza-Brito, Maurizio Piai, Alfonso Ramallo
and Martin Schvellinger. The work of E.C. and S.Y. is 
partially supported by the National Science Foundation 
under Grant No. PHY-0455649. E.C. also acknowledges support 
of CONACyT grant CB-2008-01-104649 and CONACyT's 
High Enegry Physics Network.

\appendix
\renewcommand{\thesection}{\Alph{section}}
\renewcommand{\theequation}{\Alph{section}.\arabic{equation}}
\section{Appendix: Technical aspects of the SUSY background}
\label{backgroundfunctionsappendix}
\setcounter{equation}{0}
We write in this appendix various technical aspects
of the supersymmetric backgrounds.
As explained in Section \ref{section2} one changes the basis of functions
from $\Phi, h,g,k,a,b$ into $P,Q, Y,\hat{\Phi},\tau, \sigma$ in order to 
decouple the non-linear system of BPS equations.
As explained in \cite{HoyosBadajoz:2008fw}, 
the change of basis functions is
\bea
& & 4 e^{2h}=\frac{P^2-Q^2}{P\cosh\tau -Q}, \;\; e^{2{g}}= P\cosh\tau -Q,\;\;
e^{2k}= 4 Y,\;\;\nonumber\\
& &  a=\frac{P\sinh\tau}{P\cosh\tau -Q},\;\; N_c b= \sigma.
\label{functions}
\eea
Using the relations above, one can solve for the
decoupled BPS equations, 
\bea
& & Q(\rho)=(Q_0+ N_c)\cosh\tau + N_c (2\rho \cosh\tau -1),\nonumber\\
& & \sinh\tau(\rho)=\frac{1}{\sinh(2\rho-2 \rho_0)},\quad \cosh\t(\r)=\coth(2\r-2{\r_0}),\nonumber\\
& & Y(\rho)=\frac{P'}{8},\;\;\;
e^{4\Phi}=\frac{e^{4\Phi_o} \cosh(2{\rho_0})^2}{(P^2-Q^2) Y
\sinh^2\tau},\nonumber\\
& & \sigma=\tanh\tau (Q+N_c)= \frac{(2N_c\rho + Q_o + N_c)}{\sinh(2\rho
-2{\rho_0})}.
\label{BPSeqs}
\eea
and the master equation (\ref{master}). Solving the master equation
in the UV (\ref{UV-II-N}) and plugging back into 
eqs.(\ref{functions})-(\ref{BPSeqs}) 
the background functions read at large $\r$,
\bea
& &e^{2h}\sim\Big[\frac{c_+ e^{4\r/3}}{4}+\frac{N_c}{4}(2\r-1)+\frac{N_c^2 e^{-4\r/3}}{16c_+}(16\r^2-16\r+13)+\frac{e^{-8\r/3}}{4}(c_- -c_+(2+\frac{8\r}{3}))
   \Big]\NO\\
& &\frac{e^{2g}}{4}\sim \Big[  
\frac{c_+ e^{4\r/3}}{4}-\frac{N_c}{4}(2\r-1)+\frac{N_c^2 e^{-4\r/3}}{16c_+}(16\r^2-16\r+13)+\frac{e^{-8\r/3}}{4}(c_- +c_+(2-\frac{8\r}{3})) \Big]\NO\\
& & \frac{e^{2k}}{4} \sim \Big[
\frac{c_+ e^{4\r/3}}{6}-\frac{N_c^2 e^{-4\r/3}}{24c_+}(4\r-5)^2+
\frac{e^{-8\r/3}}{3}(c_+(\frac{8\r}{3} -1) - c_-)
   \Big]\NO\\
& & e^{4\Phi - 4 \Phi_0}\sim \Big[1+\frac{3N_c^2 e^{-8\r/3}}{4c_+^2}(1-8\r)+
\frac{3 N_c^4 e^{-16\r/3}}{512 c_+^4}(2048\r^3+1152 \r^2 +2352 \r -775)   \Big]\NO\\
& &a\sim 2 e^{-2\r}+\frac{2N_c}{c_+}(2\r-1)e^{-10\r/3} +\frac{2N_c^2}{c_+^2}(2\r-1)^2 e^{-14\r/3}        \NO\\
& & b=\frac{2\r}{\sinh(2\r)}\sim 4\r e^{-\r}+ 4\r e^{-6\r}
\label{gggg}
\eea
The geometry in eq. (\ref{f3old}) asymptotes to the conifold 
after using the expansions above.
In the IR we have using eq. (\ref{P-IR}) and (\ref{functions}),
\bea
e^{2h}&\sim&\frac{h_1 \rho ^2}{2}+\frac{4}{45} \left(-6 h_1+
15 N_c-\frac{16 N_c^2}{h_1}\right) \rho ^4+\co(\r^6),\NO\\
\frac{e^{2g}}{4}&\sim&\frac{h_1}{8}+
\frac{1}{15} \left(3 h_1-5 N_c-\frac{2 N_c^2}{h_1}\right) \rho ^2+
\frac{2 \left(3 h_1^4+70 h_1^3 N_c-144 
h_1^2 N_c^2-32 N_c^4\right) \rho ^4}{1575 h_1^3}\NO\\
&&+\co(\r^6),\NO\\
\frac{e^{2k}}{4}&\sim&\frac{h_1}{8}
+\frac{\left(h_1^2-4 N_c^2\right) \rho ^2}{10 h_1}
+\frac{\left(6 h_1^4-8 h_1^2 N_c^2-64 N_c^4\right) \rho ^4}{315 h_1^3}
+\co(\r^6),\NO\\
e^{4(\Phi-\Phi_0)}&\sim&1+\frac{64  N_c^2 \rho ^2}{9 h_1^2}
+\frac{128 N_c^2 \left(-15 h_1^2+124 N_c^2\right) \rho ^4}{405 h_1^4}+
\co(\r^6),\\
a &\sim& 1+\left(-2+\frac{8 N_c}{3 h_1}\right) \rho ^2
+\frac{2 \left(75 h_1^3-232 h_1^2 N_c+160 h_1 N_c^2+64 N_c^3\right) 
\rho ^4}{45 h_1^3}
+\co(\r^6),\NO\\
b &=& \frac{2\r}{\sinh(2\r)}\sim 1-\frac{2}{3}\r^2 
+\frac{14}{45}\r^4  +\co(\r^6)  .
\label{SUSYIREXP}
\eea
This space is free of singularities as can be checked by computing invariants.

\section{Appendix: Euler-Lagrange equations of motion}
\label{eqsofmotionappendix}
\setcounter{equation}{0}
Here we write the full equations of motion, for reference.
We start with the effective Lagrangian and the constraint and then write the equations of motion. We set $N_c=1$ for simplicity.

The effective Lagrangian is $L=T-U$, with
\begin{align}
T  &=		-\frac{1}{128} e^{2 \Phi } \Bigl\{e^{4 g} \left(a'\right)^2+\left(b'\right)^2 N_c^2-8 e^{2 (g+h)} \Bigl[2 g' \left(2 h'+k'+2 \Phi '\right)+\left(g'\right)^2		\nn\\
	&\qquad\quad		+2 h' \left(k'+2 \Phi '\right)+\left(h'\right)^2+2 \Phi ' \left(k'+\Phi '\right)\Bigr]\Bigr\}   , \\ 
U	 &=		\frac{1}{256} e^{-2 (g+h-\Phi )} \Bigl[a^4 e^{4 g} \left(N_c^2+e^{4 k}\right)-4 a^3 b e^{4 g} N_c^2+2 a^2 e^{2 g} \Bigl(2 b^2 e^{2 g} N_c^2	\nn\\
	&\qquad\quad		+e^{2 g} N_c^2+4 e^{2 h} N_c^2-8 e^{2 (g+h+k)}+4 e^{4 g+2 h}-e^{2 g+4 k}+4 e^{2 h+4 k}\Bigr)		\nn\\
	&\qquad\quad		-4 a b e^{2 g} N_c^2 \left(e^{2 g}+4 e^{2 h}\right)+8 b^2 N_c^2 e^{2 (g+h)}+e^{4 g} N_c^2	+16 e^{4 h} N_c^2	\nn\\
	&\qquad\quad		-16 e^{2 (2 g+h+k)}-64 e^{2 (g+2 h+k)}+e^{4 (g+k)}+16 e^{4 (h+k)}\Bigr]	.
\end{align}
The constraint is
\begin{align}
	0     &=  T+U		\nn\\
				&=	e^{-2 (g+h-\Phi )} \Bigl[-2 \left(a'\right)^2 e^{6 g+2 h}+a^4 e^{4 g} \left(e^{4 k}+1\right)-4 a^3 b e^{4 g}\nn\\
	&\qquad\quad		+2 a^2 e^{2 g} \Bigl(2 b^2 e^{2 g}-8 e^{2 (g+h+k)}+4 e^{4 g+2 h}-e^{2 g+4 k}+e^{2 g}\nn\\
	&\qquad\quad		+4 e^{2 h+4 k}+4 e^{2 h}\Bigr)-4 a b e^{2 g} \left(e^{2 g}+4 e^{2 h}\right)-2 \left(b'\right)^2 e^{2 (g+h)}\nn\\
	&\qquad\quad		+64 e^{4 (g+h)} g' h'+32 e^{4 (g+h)} g' k'+64 e^{4 (g+h)} g' \Phi'+16 e^{4 (g+h)} \left(g'\right)^2\nn\\
	&\qquad\quad		+8 b^2 e^{2 (g+h)}+32 e^{4 (g+h)} h' k'+64 e^{4 (g+h)} h' \Phi'+16 e^{4 (g+h)} \left(h'\right)^2\nn\\
	&\qquad\quad		+32 e^{4 (g+h)} k' \Phi'-16 e^{2 (2 g+h+k)}-64 e^{2 (g+2 h+k)}+32 e^{4 (g+h)} \left(\Phi'\right)^2\nn\\
	&\qquad\quad		+e^{4 (g+k)}+e^{4 g}+16 e^{4 (h+k)}+16 e^{4 h}\Bigr].
\label{constraint}
\end{align}

The equations of motion are:
\begin{align}
	g''   &=  \frac{1}{8} e^{-4 g-2 h} \Bigl[e^{6 g} \left(a'\right)^2-4 a^2 e^{2 g+4 k}-4 a^2 e^{2 g}+4 a^2 e^{6 g}+8 a b e^{2 g}	\nn\\
	&\qquad\quad		-e^{2 g} \left(b'\right)^2-4 b^2 e^{2 g}-16 e^{4 g+2 h} g' h'-16 e^{4 g+2 h} g' \Phi'		\nn\\
	&\qquad\quad		-16 e^{4 g+2 h} \left(g'\right)^2+32 e^{2 g+2 h+2 k}-16 e^{2 h+4 k}-16 e^{2 h}\Bigr]    
\label{eq:eomg}\\
  h''   &=  -\frac{1}{8} e^{-2 g-4 h} \Bigl[\left(a'\right)^2 e^{4 g+2 h}+a^4 e^{2 g+4 k}+a^4 e^{2 g}-4 a^3 b e^{2 g}+4 a^2 b^2 e^{2 g}	\nn\\
	&\qquad\quad		-8 a^2 e^{2 g+2 h+2 k}+4 a^2 e^{4 g+2 h}-2 a^2 e^{2 g+4 k}+2 a^2 e^{2 g}	\nn\\
	&\qquad\quad		+4 a^2 e^{2 h+4 k}+4 a^2 e^{2 h}-4 a b e^{2 g}-8 a b e^{2 h}+e^{2 h} \left(b'\right)^2	\nn\\
	&\qquad\quad		+4 b^2 e^{2 h}+16 e^{2 g+4 h} g' h'+16 e^{2 g+4 h} h' \Phi'+16 e^{2 g+4 h} \left(h'\right)^2	\nn\\
	&\qquad\quad		-8 e^{2 g+2 h+2 k}+e^{2 g+4 k}+e^{2 g}\Bigr]    \\ 
	k''   &=  \frac{1}{8} e^{-4 g-4 h} \Bigl(a^4 e^{4 g+4 k}-a^4 e^{4 g}+4 a^3 b e^{4 g}-4 a^2 b^2 e^{4 g}+8 a^2 e^{2 g+2 h+4 k}	\nn\\
	&\qquad\quad		-8 a^2 e^{2 g+2 h}-8 a^2 e^{6 g+2 h}-2 a^2 e^{4 g+4 k}-2 a^2 e^{4 g}+16 a b e^{2 g+2 h}	\nn\\
	&\qquad\quad		+4 a b e^{4 g}-8 b^2 e^{2 g+2 h}-16 e^{4 g+4 h} g' k'-16 e^{4 g+4 h} h' k'	\nn\\
	&\qquad\quad		-16 e^{4 g+4 h} k' \Phi'+e^{4 g+4 k}-e^{4 g}+16 e^{4 h+4 k}-16 e^{4 h}\Bigr)     \\
	\Phi''&=  \frac{1}{8} e^{-4 g-4 h} \Bigl[a^4 e^{4 g}-4 a^3 b e^{4 g}+4 a^2 b^2 e^{4 g}+8 a^2 e^{2 g+2 h}-16 a b e^{2 g+2 h}	\nn\\
	&\qquad\quad		+2 a^2 e^{4 g}-4 a b e^{4 g}+2 \left(b'\right)^2 e^{2 g+2 h}+8 b^2 e^{2 g+2 h}-16 e^{4 g+4 h} g' \Phi'	\nn\\
	&\qquad\quad		-16 e^{4 g+4 h} h' \Phi'	-16 e^{4 g+4 h} \left(\Phi'\right)^2+e^{4 g}+16 e^{4 h}\Bigr]    \label{eq:eomphi}
\end{align}
\begin{align}  
	a''   &=  e^{-4 g-2 h} \Bigl(-4 a' e^{4 g+2 h} g'-2 a' e^{4 g+2 h} \Phi'+a^3 e^{2 g+4 k}+a^3 e^{2 g}-3 a^2 b e^{2 g}	\nn\\
	&\qquad\quad		+2 a b^2 e^{2 g}-8 a e^{2 g+2 h+2 k}+4 a e^{4 g+2 h}-a e^{2 g+4 k}+a e^{2 g}	\nn\\
	&\qquad\quad		+4 a e^{2 h+4 k}+4 a e^{2 h}-b e^{2 g}-4 b e^{2 h}\Bigr)    \\ 
	b''   &=  -e^{-2 h} \left(a^3 e^{2 g}-2 a^2 b e^{2 g}+a e^{2 g}+4 a e^{2 h}+2 e^{2 h} b' \Phi'-4 b e^{2 h}\right)	\label{eq:eomb}
\end{align}

\section{Appendix: Explicit expansion of the functions}
\label{explicitexpansionsappendix}
\setcounter{equation}{0}
Here we include the explicit solutions for the expansions \eqref{UVexpansionnonsusy} and \eqref{IRnonsusyexpansion}. In this section we again set $N_c=1$.

\subsection{UV} 
	
\begin{align}
	e^{2 g}   &=   c_+ e^{4 \rho /3}  -\left(2 c_+ W_{20}^2+4 H_{11} \rho +Q_o\right)	\nn\\
	&\quad	 -\frac{1}{48 c_+}\Bigl\{-12 H_{11} \left[(32 \rho -6) Q_o-c_+ W_{20}^2 (8 \rho +93)\right]-12 c_+ W_{20}^2 Q_o	\nn\\
	&\quad  +72 c_+^2 W_{20} e^{2 \rho _o}+120 c_+^2 W_{20}^4 \rho -26 c_+^2 W_{20}^4+12 c_+^2 \Phi _{30} e^{-4 \Phi_\infty} -72 \rho	\nn\\
	&\quad  -24 H_{11}^2 \left(32 \rho ^2-12 \rho +15\right)-48 Q_o^2+9\Bigr\}e^{-4 \rho /3} 
 								 +O\bigl(e^{-8 \rho /3}\bigr)   \label{UVexpg} \\
	e^{2 h}   &=   \frac{c_+}{4} e^{4 \rho /3}+\left(H_{11} \rho +\frac{Q_o}{4}\right)	\nn\\
	&\quad	 -\frac{1}{192 c_+}\Bigl\{-12 H_{11} \left[c_+ W_{20}^2 (88 \rho +75)+(32 \rho -6) Q_o\right]-396 c_+ W_{20}^2 Q_o	\nn\\
	&\quad	 +264 c_+^2 W_{20} e^{2 \rho _o}+440 c_+^2 W_{20}^4 \rho -626 c_+^2 W_{20}^4 +12 c_+^2 \Phi _{30} e^{-4 \Phi_\infty}-72 \rho 	\nn\\
	&\quad	 -24 H_{11}^2 \left(32 \rho ^2-12 \rho +15\right)-48 Q_o^2+9\Bigr\}e^{-4 \rho /3} +O\bigl(e^{-8 \rho /3}\bigr)    
\\ 
	e^{2 k}   &=   \frac{2 c_+}{3 }e^{4 \rho /3} +\frac{c_+ W_{20}^2}{3} 
								-\frac{1}{24 c_+}\Bigl[4 H_{11} (16 \rho -9) \left(3 c_+ W_{20}^2+2 Q_o\right) +16 Q_o^2	\nn\\
	&\quad	 +84 c_+ W_{20}^2 Q_o-72 c_+^2 W_{20} e^{2 \rho _o}-120 c_+^2 W_{20}^4 \rho +190 c_+^2 W_{20}^4-24 \rho -9	\nn\\
	&\quad	 +4 c_+^2 \Phi _{30} e^{-4 \Phi_\infty}+8 H_{11}^2 \left(32 \rho ^2-36 \rho +27\right)\Bigr]e^{-4 \rho /3}
						+O\bigl(e^{-8 \rho /3}\bigr)    
\\ 
	e^{4\Phi} &=   e^{4 \Phi_\infty}
								+\left(\Phi _{30}-\frac{6 \rho  e^{4 \Phi_\infty}}{c_+^2}\right) e^{-8 \rho /3}	\nn\\
	&\quad	 +\frac{W_{20}^2 \left(32 c_+^2 \Phi _{30}-3 (64 \rho +25) e^{4 \Phi_\infty}\right)}{24 c_+^2}e^{-4\rho}
					 +O\bigl(e^{-16 \rho /3}\bigr)	
\end{align}
\begin{align}
a   &=   W_{20} e^{-2 \rho /3}
					+\left(\frac{4 H_{11} W_{20} \rho }{c_+}+2 e^{2 \rho _o}+\frac{10 W_{20}^3 \rho }{3}\right) e^{-2 \rho} 	\nn\\
	&\quad	+\frac{1}{48 c_+^2}\Biggl(4 c_+ \biggl\{H_{11} \left[96 \rho  e^{2 \rho _o}+W_{20}^3 \left(160 \rho ^2+552 \rho +495\right)\right] +2 Q_o \Bigl[12 e^{2 \rho _o}	\nn\\
	&\quad	+W_{20}^3 (20 \rho +87)\Bigr]\biggr\}+c_+^2 \left[W_{20}^5 (391-480 \rho )-288 W_{20}^2 e^{2 \rho _o}\right]	\nn\\
	&\quad	+12 W_{20} \Bigl[2 H_{11} (40 \rho +27) Q_o	\nn\\
	&\quad	+6 H_{11}^2 \left(32 \rho ^2+36 \rho +43\right)+8 Q_o^2-15\Bigr]\Biggr)e^{-10 \rho /3}	+O\bigl(e^{-4\rho}\bigr)
\\ 
b   &=   \frac{9}{4} W_{20} e^{-2 \rho /3}
						+\left\{\left[4 e^{2 \rho _o}+\frac{W_{20}^3}{6} (20 \rho -23)-\frac{12 W_{20} Q_o}{6 c_+}\right]\rho+V_{40}\right\} e^{-2 \rho } 	\nn\\
	&\quad		+\frac{ 3 W_{20}  e^{-4 \Phi_\infty} }{512 c_+^2} \biggl\{e^{4 \Phi_\infty} \Bigl[-3456 H_{11} \left(3 c_+ W_{20}^2+2 Q_o\right)-3240 c_+ W_{20}^2 Q_o 	\nn\\
	&\quad		+192 c_+^2 V_{40} W_{20}+2819 c_+^2 W_{20}^4-35712 H_{11}^2-576 Q_o^2+882\Bigr] 	\nn\\
	&\quad		-16 \rho  e^{4 \Phi_\infty} \Bigl[144 H_{11} \left(3 c_+ W_{20}^2+2 Q_o\right)+24 c_+ W_{20}^2 Q_o 	\nn\\
	&\quad		-c_+^2 W_{20} \left(48 e^{2 \rho _o}+269 W_{20}^3\right)+1728 H_{11}^2-18\Bigr]+3024 c_+^2 W_{20} e^{2 \rho _o+4 \Phi_\infty} 	\nn\\
	&\quad		-128 \rho ^2 e^{4 \Phi_\infty} \left(72 H_{11}^2-5 c_+^2 W_{20}^4\right)-48 c_+^2 \Phi _{30}\biggr\}e^{-10 \rho /3}
					+O\bigl(e^{-4 \rho}\bigr)
	\label{UVexpb}
\end{align}

We can see the effect of the SUSY-breaking parameters by looking at functions 
of the form $\Delta(e^{2g})= e^{2g} - e^{2g_\text{SUSY}}$, where $g$ corresponds to the full solution and $g_\text{SUSY}$ corresponds to the SUSY case with $Q_o=-N_c$ and $\rho_o=0$. Note that in general only one of the two solutions will go to the regular IR --- if we start with a SUSY solution and turn on one of the SUSY-breaking parameters while keeping e.g. $c_+$ fixed, we will have to change $c_-$ to recover the regular IR. Those SUSY-breaking parameters which have non-zero values in the SUSY case are expressed here in terms of e.g. $\Delta H_{11} = H_{11} - H_{11}^{\text{SUSY}}$.

\begin{align}
	\Delta \bigl(e^{2 g}\bigr)   
			&=   \left(-2 c_+ W_{20}^2-\Delta Q_o-4 \Delta H_{11} \rho \right) 	\nn\\
	&\quad		+\frac{e^{-4 \Phi_\infty} }{24 c_+}\Biggl\{-3 c_+ W_{20}^2 e^{4 \Phi_\infty} \Bigl[-2 \Delta Q_o+2 \Delta H_{11} (8 \rho +93)+8 \rho +95\Bigr] 	\nn\\
	&\quad		-c_+^2 \left[36 W_{20} e^{2 \rho _o+4 \Phi_\infty}+W_{20}^4 (60 \rho -13) e^{4 \Phi_\infty}+6 \Delta \Phi _{30}\right] 	\nn\\
	&\quad		+6 e^{4 \Phi_\infty} \biggl[\Delta H_{11} \Bigl((32 \rho -6) \Delta Q_o+64 \rho ^2-56 \rho +36\Bigr) +\Delta Q_o \left(4 \Delta Q_o+16 \rho -11\right) 	\nn\\
	&\quad		+\Delta H_{11}^2 \left(64 \rho ^2-24 \rho +30\right)\biggr]\Biggr\}e^{-4 \rho /3} +O\bigl(e^{-8 \rho /3}\bigr)   
\\
\Delta \bigl(e^{2 h}\bigr) 
			&=   \left(\frac{\Delta Q_o}{4}+\Delta H_{11} \rho \right)+\frac{e^{-4 \Phi_\infty}}{96 c_+}  \Biggl(3 c_+ W_{20}^2 e^{4 \Phi_\infty} \Bigl[66 \Delta Q_o+2 \Delta H_{11} (88 \rho +75)+88 \rho +9\Bigr] 	\nn\\
	&\quad		-c_+^2 \left[132 W_{20} e^{2 \rho _o+4 \Phi_\infty}+W_{20}^4 (220 \rho -313) e^{4 \Phi_\infty} 	+6 \Delta \Phi _{30}\right] 	\nn\\
	&\quad		+6 e^{4 \Phi_\infty} \biggl\{\Delta H_{11} \left[(32 \rho -6) \Delta Q_o+64 \rho ^2-56 \rho +36\right]
						+\Delta Q_o \left(4 \Delta Q_o+16 \rho -11\right) 	\nn\\
	&\quad		+\Delta H_{11}^2 \left(64 \rho ^2-24 \rho +30\right)\biggr\}\Biggr)e^{-4 \rho /3}	+O\bigl(e^{-8 \rho /3}\bigr)		
\\
\Delta \bigl(e^{2 k}\bigr)   
			&=   \frac{1}{3} c_+ W_{20}^2+\frac{e^{-4 \Phi_\infty}}{12 c_+}\Biggl(-3 c_+ W_{20}^2 e^{4 \Phi_\infty} \Bigl[14 \Delta Q_o+2 \Delta H_{11} (16 \rho -9)+16 \rho -23\Bigr] 	\nn\\
	&\quad		+c_+^2 \left[36 W_{20} e^{2 \rho _o+4 \Phi_\infty}+5 W_{20}^4 (12 \rho -19) e^{4 \Phi_\infty}-2 \Delta \Phi _{30}\right] 	\nn\\
	&\quad		-2 e^{4 \Phi_\infty} \Bigl\{2 \Delta H_{11} \left[(16 \rho -9) \Delta Q_o+4 \left(8 \rho ^2-13 \rho +9\right)\right] 	\nn\\
	&\quad		+\Delta Q_o \left(4 \Delta Q_o+16 \rho -17\right) 	\nn\\
	&\quad		+\Delta H_{11}^2 \left(64 \rho ^2-72 \rho +54\right)\Bigr\}\Biggr)e^{-4 \rho /3}
						+O\bigl(e^{-8 \rho /3}\bigr)    
\\
\Delta \bigl(e^{4 \Phi }\bigr)   
			&=   \Delta \Phi _{30} e^{-8 \rho /3}+\frac{W_{20}^2}{24 c_+^2}  \left[32 c_+^2 \Delta \Phi _{30}-3 (64 \rho +17) e^{4 \Phi_\infty}\right]e^{-4 \rho}+O\bigl(e^{-16 \rho /3}\bigr)   
\end{align}
\begin{align}
\Delta a
				&=   W_{20} e^{-2 \rho /3}
						+\frac{1}{c_+}\left[\frac{2}{3} c_+ \left(3 e^{2 \rho _o}+5 W_{20}^3 \rho -3\right)+2 W_{20} \left(2 \Delta H_{11}+1\right) \rho \right]e^{-2 \rho}  	\nn\\
	&\quad		+\frac{1}{48 c_+^2} \Biggl(2 c_+ \biggl\{  W_{20}^3 \Bigl[4 (20 \rho +87) \Delta Q_o+2 \Delta H_{11} \left(160 \rho ^2+552 \rho +495\right) 	\nn\\
	&\quad		+160 \rho ^2+472 \rho +147\Bigr]+48 \left[4 \Delta H_{11} \rho  e^{2 \rho _o}+\Delta Q_o e^{2 \rho _o}+(2 \rho -1) \left(e^{2 \rho _o}-1\right)\right]  \biggr\} 	\nn\\
	&\quad		+c_+^2 \left[W_{20}^5 (391-480 \rho )-288 W_{20}^2 e^{2 \rho _o}\right] 
						+6 W_{20} \biggl\{+(80 \rho +22) \Delta Q_o 	\nn\\
	&\quad		+4 \Delta H_{11} \left[(40 \rho +27) \Delta Q_o+96 \rho ^2+68 \rho +102\right]
						+16 \Delta Q_o^2+12 \Delta H_{11}^2 \left(32 \rho ^2+36 \rho +43\right) 	\nn\\
	&\quad		+96 \rho ^2+28 \rho +61\biggr\}\Biggr)e^{-10 \rho /3}
						+O\bigl(e^{-14 \rho /3}\bigr)    \\ 
\Delta b  
				&=   \frac{9}{4} W_{20} e^{-2 \rho /3}
						+ \Bigl[c_+ \left\{\rho  \left[24 \left(e^{2 \rho _o}-1\right)+W_{20}^3 (20 \rho -23)\right]+6 V_{40}\right\} 	\nn\\
	&\quad		-12 W_{20} \rho  \left(\Delta Q_o-1\right)\Bigr]\frac{e^{-2 \rho}}{6 c_+}
						+\frac{3 W_{20}  e^{-4 \Phi_\infty}}{512 c_+^2} \Biggl(c_+^2 \biggl\{48 W_{20} e^{4 \Phi_\infty} \left[(16 \rho +63) e^{2 \rho _o}+4 V_{40}\right] 	\nn\\
	&\quad		+W_{20}^4 \left(640 \rho ^2+4304 \rho +2819\right) e^{4 \Phi_\infty}-48 \Delta \Phi _{30}\biggr\} 	\nn\\
	&\quad		-24 c_+ W_{20}^2 e^{4 \Phi_\infty} \left[(16 \rho +135) \Delta Q_o+144 \Delta H_{11} (2 \rho +3)+128 \rho +81\right] 	\nn\\
	&\quad		-18 e^{4 \Phi_\infty} \biggl\{64 \Delta H_{11} \left[(4 \rho +6) \Delta Q_o+8 \rho ^2+20 \rho +25\right] 	\nn\\
	&\quad		+128 (\rho +1) \Delta Q_o+32 \Delta Q_o^2+64 \Delta H_{11}^2 \left(8 \rho ^2+24 \rho +31\right) 	\nn\\
	&\quad		+128 \rho ^2+240 \rho +289\biggr\}\Biggr)e^{-10 \rho /3}
						+O\bigl(e^{-14 \rho /3}\bigr)    
\end{align}

\subsection{IR}

\begin{align}
	e^{2 g}   &=   \frac{h_1}{2}-\frac{1}{8} \left[4k_2-h_1 \left(w_2^2+4\right)+\frac{4}{h_1} \left(v_2^2+4\right)\right]\rho^2
								+ \frac{1}{20160 h_1^3}\biggl\{1600 h_1 k_2 \left(3 v_2^2+8\right)			\nn\\
	&\qquad				-8 h_1^2 \Bigl[210 k_2^2+144 v_2 \left(w_2-3\right) w_2+3 v_2^2 \left(105 w_2^2+168 w_2+580\right)		\nn\\
	&\qquad				+4 \left(55 w_2^2+360 w_2+652\right)\Bigr]-480 h_1^3 k_2 \left(w_2^2+18 w_2+18\right)						\nn\\
	&\qquad				+3 h_1^4 \left(75 w_2^4+432 w_2^3+488 w_2^2+960 w_2+1520\right)					\nn\\
	&\qquad				+16 \left(405 v_2^4+1592 v_2^2+656\right)\biggr\}\rho ^4+O\left(\rho ^6\right)    
\label{IRexpg}
\end{align}
\begin{align}
e^{2 h}   &=   \frac{h_1}{2}\rho ^2+\frac{1}{72}\left[24 k_2-3 h_1\left(3 w_2^2+4\right)-\frac{4}{h_1}(9 v_2^2-4)\right]\rho ^4 	\nn\\
	&\qquad				+\frac{1}{907200 h_1^3} \biggl\{24 h_1^2 \Bigl[5250 k_2^2+432 v_2 w_2 \left(23 w_2+57\right)		\nn\\
	&\qquad				-9 v_2^2 \left(189 w_2^2+2016 w_2+940\right)-20940 w_2^2+8640 w_2+5680\Bigr]				\nn\\
	&\qquad				+480 h_1 k_2 \left(9 v_2^2-172\right)-360 h_1^3 k_2 \left(627 w_2^2-432 w_2+44\right)			\nn\\
	&\qquad				+9 h_1^4 \left(2007 w_2^4-6624 w_2^3+11352 w_2^2-5760 w_2+1520\right)			\nn\\
	&\qquad				+16 \left(19359 v_2^4+27288 v_2^2-22672\right)\biggr\}\rho ^6+O\left(\rho ^8\right)
\\
e^{2 k}   &=   \frac{h_1}{2}+k_2 \rho ^2+\frac{2}{315 h_1^3}  \biggl\{5 h_1 k_2 \left(27 v_2^2+2\right)
								+\frac{15}{4} h_1^3 k_2 \left(9 w_2^2+36 w_2+22\right)		\nn\\
	&\qquad				+\frac{3}{2} h_1^2 \Bigl[175 k_2^2+12 v_2 w_2 \left(w_2+4\right)-30 v_2^2+30 w_2^2+120 w_2+208\Bigr]			\nn\\
	&\qquad				-\frac{9}{16} h_1^4 \left(w_2^4+8 w_2^3+36 w_2^2+80 w_2+80\right)		\nn\\
	&\qquad				+9 v_2^4+36 v_2^2-528\biggr\}\rho ^4+O\left(\rho ^6\right)
\\
e^{4 \Phi}&=   e^{4 \phi _0}+\frac{4e^{4 \phi _0}}{3 h_1^2} \left(3 v_2^2+4\right) \rho ^2
								+\frac{e^{4 \phi _0}}{135 h_1^4} \biggl\{-60 h_1 k_2 \left(3 v_2^2-8\right)		\nn\\
	&\qquad				+3 h_1^2 \Bigl[3 v_2^2 \left(9 w_2^2+36 w_2+40\right)	-36 v_2 w_2 \left(w_2+4\right)-176\Bigr]	\nn\\
	&\qquad				+4 \left(243 v_2^4+672 v_2^2+944\right)\biggr\}\rho ^4 +O\left(\rho ^6\right)
\\
a   			&=   1+w_2 \rho ^2+\frac{1}{90 h_1^2} \biggl\{w_2 \Bigl[150 h_1 k_2-3 h_1^2 \left(6 w_2^2-9 w_2+28\right)+400\Bigr]	\nn\\
	&\qquad				+36 v_2^2 \left(w_2+2\right)-72 v_2 \left(w_2+2\right)\biggr\}\rho ^4+O\left(\rho ^6\right)
\end{align}
\begin{align} 
b   			&=   1+v_2 \rho ^2-\frac{1}{90 h_1^2} \biggl\{v_2 \Bigl[30 h_1 k_2-h_1^2 \left(9 w_2^2+36 w_2+60\right)+176\Bigr]	\nn\\
	&\qquad				+9 h_1^2 w_2 \left(w_2+4\right)+72 v_2^3\biggr\}\rho ^4+O\left(\rho ^6\right)    
\end{align}

The effect of the SUSY-breaking parameters can be seen from the differences:

\begin{align}
\Delta \bigl(e^{2 g}\bigr)   &=   \frac{1}{24 h_1}\Bigl[-4 h_1 \left(3 \Delta k_2-4 \Delta w_2\right)+3 h_1^2 \left(\Delta w_2-4\right) \Delta w_2+4 \left(4-3 \Delta v_2\right) \Delta v_2\Bigr]\rho ^2 		\nn\\
		&\qquad					+\frac{1}{60480 h_1^3}\biggl\{-24 h_1^2 \Bigl[320 \Delta k_2 \left(\Delta w_2+7\right)+210 \Delta k_2^2 -4 \Delta v_2 \left(69 \Delta w_2^2+224\right)		\nn\\
		&\qquad					+3 \Delta v_2^2 \left(105 \Delta w_2^2-252 \Delta w_2+584\right)+8 \left(112-129 \Delta w_2\right) \Delta w_2\Bigr]	\nn\\
		&\qquad					+16 \Delta v_2 \left(1215 \Delta v_2^3-3240 \Delta v_2^2+3216 \Delta v_2-2944\right)		\nn\\
		&\qquad					+9 h_1^4 \Delta w_2 \left(75 \Delta w_2^3-168 \Delta w_2^2-368 \Delta w_2+896\right)	\nn\\
		&\qquad					-96 h_1^3 \Bigl[3 \Delta k_2 \left(5 \Delta w_2^2+70 \Delta w_2-56\right)+\Delta w_2 \left(-75 \Delta w_2^2+126 \Delta w_2+184\right)\Bigr]		\nn\\
		&\qquad					+64 h_1 \Bigl[3 \Delta k_2 \left(75 \Delta v_2^2-100 \Delta v_2+264\right) -126 \Delta v_2^2 \left(5 \Delta w_2-6\right)		\nn\\
		&\qquad					+552 \Delta v_2 \Delta w_2+464 \Delta w_2\Bigr]     \biggr\}\rho ^4 +O\left(\rho ^6\right)
\\
\Delta\bigl(e^{2 h}\bigr)   &=   \frac{1}{24 h_1}\Bigl[8 h_1 \left(\Delta k_2-2 \Delta w_2\right)-3 h_1^2 \left(\Delta w_2-4\right) \Delta w_2+4 \left(4-3 \Delta v_2\right) \Delta v_2\Bigr]\rho ^4 		\nn\\
		&\qquad					+\frac{1}{302400 h_1^3}\biggl\{24 h_1^2 \Bigl[-80 \Delta k_2 \left(209 \Delta w_2-490\right)+1750 \Delta k_2^2		\nn\\
		&\qquad					+\Delta v_2^2 \left(-567 \Delta w_2^2-3780 \Delta w_2+7032\right)+4 \Delta v_2 \left(1017 \Delta w_2^2-3136\right)	\nn\\
		&\qquad					+40 \Delta w_2 \left(157 \Delta w_2-1288\right)\Bigr]
										+32 h_1 \Bigl[5 \Delta k_2 \left(9 \Delta v_2^2-12 \Delta v_2-4352\right)	\nn\\
		&\qquad					-4 \left(189 \Delta v_2^2 \left(3 \Delta w_2+10\right)-4068 \Delta v_2 \Delta w_2+856 \Delta w_2\right)\Bigr]	\nn\\
		&\qquad					-24 h_1^3 \Bigl[5 \Delta k_2 \left(627 \Delta w_2^2-2940 \Delta w_2+3136\right)		\nn\\
		&\qquad					-4 \Delta w_2 \left(669 \Delta w_2^2-5670 \Delta w_2+14872\right)\Bigr]				\nn\\
		&\qquad					+9 h_1^4 \Delta w_2 \left(669 \Delta w_2^3-7560 \Delta w_2^2+29744 \Delta w_2-49280\right)		\nn\\
		&\qquad					+48 \Delta v_2 \left(2151 \Delta v_2^3-5736 \Delta v_2^2+6704 \Delta v_2+7936\right)\biggr\}\rho ^6 +O\left(\rho ^8\right)
\end{align}
\begin{align} 
\Delta \bigl(e^{2 k}\bigr)   &=   \Delta k_2 \rho ^2
										+\frac{1}{7560 h_1^3} \biggl\{36 h_1^3 \Bigl[15 \Delta k_2 \left(3 \Delta w_2^2+14\right)-8 \Delta w_2 \left(\Delta w_2^2-6\right)\Bigr]		\nn\\
		&\qquad					+72 h_1^2 \left(120 \Delta k_2 \Delta w_2+175 \Delta k_2^2+12 \Delta v_2 \Delta w_2^2+6 \Delta v_2^2-56 \Delta v_2-30 \Delta w_2^2\right)		\nn\\
		&\qquad					+16 h_1 \Bigl[15 \Delta k_2 \left(27 \Delta v_2^2-36 \Delta v_2-106\right)+16 \left(18 \Delta v_2-29\right) \Delta w_2\Bigr]\nn\\
		&\qquad					+16 \Delta v_2 \left(27 \Delta v_2^3-72 \Delta v_2^2-468 \Delta v_2+1072\right)		\nn\\
		&\qquad					-27 h_1^4 \Delta w_2^2 \left(\Delta w_2^2-12\right)\biggr\}\rho ^4+O\left(\rho ^6\right)    
\\
\Delta \bigl(e^{4 \Phi }\bigr)   &=   \frac{4 e^{4 \phi _0}\Delta v_2}{3 h_1^2} \left(3 \Delta v_2-4\right)  \rho ^2
										+\frac{ e^{4 \phi _0}}{135 h_1^4} \biggl\{4 h_1 \Bigl[\Delta k_2 \left(-45 \Delta v_2^2+60 \Delta v_2+100\right)	\nn\\
		&\qquad					+36 \left(3 \Delta v_2^2-8 \Delta v_2+4\right) \Delta w_2\Bigr]
										+3 h_1^2 \Bigl[3 \Delta v_2^2 \left(9 \Delta w_2^2-4\right) \nn\\
		&\qquad					-8 \Delta v_2 \left(9 \Delta w_2^2-20\right)+36 \Delta w_2^2\Bigr]	
										+4 \Delta v_2 \bigl(243 \Delta v_2^3-648 \Delta v_2^2	\nn\\
		&\qquad					+1536 \Delta v_2-1664\bigr)\biggr\}\rho ^4+O\left(\rho ^6\right)  
\\
\Delta a   &=   \Delta w_2 \rho ^2
										+\frac{1}{90 h_1^3} \biggl\{	4 h_1 \Bigl[100 \Delta k_2+\left(9 \Delta v_2^2-30 \Delta v_2-40\right) \Delta w_2\Bigr]		\nn\\
		&\qquad					+6 h_1^2 \Bigl[25 \Delta k_2 \left(\Delta w_2-2\right)-24 \left(\Delta w_2-5\right) \Delta w_2\Bigr]
										+32 \Delta v_2 \left(3 \Delta v_2-10\right)		\nn\\
		&\qquad					-3 h_1^3 \Delta w_2 \left(6 \Delta w_2^2-45 \Delta w_2+116\right)\biggr\}\rho ^4+O\left(\rho ^6\right)
\\
\Delta b   &=   \Delta v_2 \rho ^2
								+\frac{1}{90 h_1^2}\biggl\{h_1 \Bigl[\Delta k_2 \left(20-30 \Delta v_2\right)+16 \left(3 \Delta v_2-5\right) \Delta w_2\Bigr] \nn\\
		&\qquad			+3 h_1^2 \Bigl[\Delta v_2 \left(3 \Delta w_2^2+4\right)-5 \Delta w_2^2\Bigr]		\nn\\
		&\qquad			-8 \Delta v_2 \left(9 \Delta v_2^2-18 \Delta v_2+20\right)\biggr\}\rho ^4 +O\left(\rho ^6\right) 
\label{IRexpb}
\end{align}

\section{Appendix: Details of the numerical analysis}\label{appendixdetailsnum}
\setcounter{equation}{0}
\label{NumApp}
Here we shall discuss in more detail our approach to connecting the given IR and UV expansions numerically.  We start by noting that we have chosen to solve the equations of motion (\ref{eq:eomg}--\ref{eq:eomb}) starting from the IR boundary conditions. As the IR parameter space is much smaller than that of the UV, this makes a search for solutions with the correct behaviour much less computationally expensive than if we started from the UV. 

We use as our boundary conditions the IR expansions (\ref{IRexpg}--\ref{IRexpb}), extended up to order $\rho^8$. Using {\ttfamily NDSolve} in Mathematica 7 we then are able to generate numerical solutions which extend into the UV.  We start at $\rho_{IR}=10^{-4}$ as we found that in the SUSY case this gives approximately optimal accuracy. We use 40-digit {\ttfamily WorkingPrecision} in {\ttfamily NDSolve}.

Comparing the numerical solutions obtained by this method, with the known behaviour in the SUSY case, suggests that the results are trustable up to $\rho\sim11$. This is supported by the observation that the constraint \eqref{constraint} is almost
completely satisfied over this range. 
More explicitly, we find $T+U \lesssim 10^{-8}$ throughout this range. In fact it appears that the numerical solutions only fail when $b$ decreases past $\sim 10^{-9}$. In the SUSY case (in which $b\sim e^{-2\rho}$) this does correspond to $\rho\sim11$, but in the non-SUSY case (with $b\sim e^{2\rho/3})$ it occurs further into the UV.


In the IR we have five parameters $\{h_{1},\phi_{0},w_{2},k_{2},v_{2}\}$ which we can manipulate although we set $\phi_0=0$ (along with $N_c=1$) without loss of generality.
Given a value of $h_1$ we want to study the effects of the SUSY-breaking deformations for the remaining three $\{w_{2},k_{2},v_{2}\}$. We find that for a general deformation of these IR parameters the resulting solution does not exhibit the expected UV behaviour. Initially we find that the general behaviour of solutions in this parameter space has
\begin{align}
b\sim \pm e^{2\rho}\quad \text{and}\quad e^{2g}\sim e^{2h}\sim e^{2k}\sim  e^{8\r/3}
\end{align}
going into the UV. The $e^{8\r/3}$ behaviour appears to be suppressed by a very small numerical factor relative to the expected $e^{4\rho/3}$ term, and in fact is not visible in plots of $g$, $h$ and $k$ themselves. It is apparent, however, if we examine quantities of the form
\begin{align}
	e^{2k} - e^{2k_\text{SUSY}} \sim e^{8\rho/3},
\end{align}
in which the $e^{4\rho/3}$ behaviour (almost) cancels.


Given a value for one of the three non-SUSY deformations, we believe it is possible to obtain the desired UV behaviour (\ref{UVexpg}--\ref{UVexpb}), i.e.
\begin{align}
					 e^{-2\rho/3}\quad \text{and}\quad e^{2g}\sim e^{2h}\sim e^{2k}\sim  e^{4\r/3},
\end{align}
with the correct choice of the remaining two. In practice it seems easier to vary three but keep one very close to its starting value.

Having obtained a numerical solution with the correct UV behaviour, we look to determine the corresponding values of the expansion coefficients in the UV, i.e. 
\begin{align}
\{c_{+},c_{-},\Phi_\infty, Q_o, \rho_o, H_{11}, W_{20}, \Phi_{30}, V_{40}\}.
\end{align}
We define the mismatch function
\begin{align}
					m = \sum_i \left[ f_i^\text{Numerical}(\rho_\text{match}) - 
														f_i^\text{Expansion}(\rho_\text{match}) \right]^2,
\end{align}
with $f_i \in \{ g,h,k,\Phi,a,b,g',h',k',\Phi',a',b' \}$. We then minimise $m$ to match our numerical solution and a UV expansion using {\ttfamily NMinimize} (with 60-digit {\ttfamily WorkingPrecision})  at a large $\rho$ value, $\rho_{match}$.


With this setup and given the SUSY IR, {\ttfamily NMinimize} recovers the SUSY values for the UV parameters with an acceptable accuracy, even allowing all nine parameters to vary. The only restrictions we apply to the parameter space are $c_{+}\geq 0$ and $\Phi_\infty\geq\phi_0=0$.

We now present a non-SUSY solution found using the above methods for one set of values of the IR parameters. It has the expected behaviour for all functions at least up to $\rho\sim11$ (where the corresponding SUSY solution fails) and possibly as far as $\rho\sim17$. We first choose $h_1=5$ (and have set $\phi_{0}=0$ as mentioned above). The corresponding SUSY solution has
\begin{align}
	w_2		=		\frac{8}{3h_1}-2 				= -\frac{22}{15},		\qquad
	k_2		=		\frac{2}{5h_1}(h_1^2-4)	= \frac{42}{25},		\qquad
	v_2		=		-\frac23. \nn
\end{align}
This results in an {\ttfamily NMinimize} output (with $\rho_{match}=6$) of
\begin{center}
\begin{tabular}{r @{\ } c @{\ } l		@{\quad} 	r @{\ } c @{\ } l		@{\quad} 	r @{\ } c @{\ } l}
					$c_+	 				$	&	$\approx$	&	$\phantom{-}1.6							$,	&			
					$c_-					$	&	$\approx$	&	$\phantom{-}2.0 \times 10^3	$,	&		
					$\Phi_\infty	$	&	$\approx$	&	$0.076						$,	\\
					$Q_o					$	&	$\approx$	&	$-1.0	$,	&		
					$\r_o					$	&	$\approx$	&	$-6.8 \times 10^{-11}	$,	&		
					$W_{20}				$	&	$\approx$	&	$ 6.9 \times 10^{-14}	$,	\\
					$V_{40}				$	&	$\approx$	&	$\phantom{-} 2.7 \times 10^{-9}		$,	&		
					$H_{11}				$	&	$\approx$	&	$\phantom{-}0.50	$,	&		
					$\Phi_{30}		$	&	$\approx$	&	$0.38	$.
\end{tabular}
\end{center}
The associated mismatch value is $m\lesssim 10^{-29}$. We take this as a good value for the mismatch as we know that the SUSY solution does indeed exist, and these values are in good agreement with \eqref{recoverSUSYUV}.

To obtain a non-SUSY deformation, we follow the procedure described and modify the three IR parameters $\{k_2,v_2,w_2\}$ away from their SUSY values so as to manually scan the parameter space,
until we gain a solution with the correct UV behaviour. We find that a suitable choice of deformations is\footnote{
The exact values used were $\Delta k_2 = -24\,705\,875 \times 10^{-12}$,\\ $\Delta v_2 = 25\,744\,091\,286\,331\,971\,640\,358 \times 10^{-27}$ and $\Delta w_2 = 1\,029\,383\,373\,181\,636\,875 \times 10^{-22}.$
}
\begin{align}
					\Delta k_2 \approx -2.471 \times 10^{-5}	,	\quad
					\Delta v_2 \approx 2.574 \times 10^{-5}	, \quad
					\Delta w_2 \approx 1.029 \times 10^{-4} .	\nn
\end{align}
The minimization routine (again at $\rho_{match}=6$) then finds that the UV parameters are modified from their SUSY values according to
\begin{center}
\begin{tabular}{r @{\ } c @{\ } l		@{\quad} 	r @{\ } c @{\ } l		@{\quad} 	r @{\ } c @{\ } l}
					$\Delta c_+ 				$	&	$\approx$	&	$-						6.6 \times 10^{-6}	$,	&			
					$\Delta c_-					$	&	$\approx$	&	$\phantom{-} 	1.6 								$,	&		
					$\Delta \Phi_\infty	$	&	$\approx$	&	$-						4.0 \times 10^{-7}	$,	\\
					$\Delta Q_o					$	&	$\approx$	&	$-						1.5 \times 10^{-4}		$,	&		
					$\Delta \r_o				$	&	$\approx$	&	$-						7.1 \times 10^{-5}		$,	&		
					$\Delta W_{20}			$	&	$\approx$	&	$\phantom{-} 	5.2 \times 10^{-5}		$,	\\
					$\Delta V_{40}			$	&	$\approx$	&	$\phantom{-} 	5.6 \times 10^{-4}		$,	&		
					$\Delta H_{11}			$	&	$\approx$	&	$\phantom{-} 	9.1 \times 10^{-5}		$,	&		
					$\Delta \Phi_{30}		$	&	$\approx$	&	$-						5.0 \times 10^{-5}$,
\end{tabular}
\end{center}
again with a mismatch value of $m \lesssim 10^{-29}$. However, we are unsure of the precision of these values --- they appear to be slightly sensitive to the value of $\rho_{match}$, and so should be interpreted with caution. We present plots of the functions 
(figure \ref{fig:functionPlots}) in the main text.


\section{Appendix: Free Energy }
\setcounter{equation}{0}
\label{Freeenergyappendix}

Consider the Euclidean action ${\cal I}$  for the wrapped D5 background of section \eqref{sec:susybreaking}. The free energy is 
$F={\cal I}/\beta $, where $\beta$ is the period of the compactified time direction.  
\begin{multline}
	\label{eq:I-nonreg}
{\cal I} = {\cal I}_{\rm grav}	+{\cal I}_{\rm surf}\\
=-\frac{1}{16 \pi } \int_{\cal M} d^{10}x \sqrt{\bf{g}}  R  +
\frac{1}{32\pi}\int_{\cal M}\left( 
d\Phi \wedge \star  d\Phi  + e^{\Phi}
F_{3}\wedge\star F_{3}\right)\\
  - \frac{1}{8\pi }\oint_{\Sigma_r} { ^9 K d\Sigma_r}.  
\end{multline}
 ${\cal M}$ is a ten dimensional volume
enclosed by a nine dimensional 
boundary $\Sigma_r$. The boundary $\Sigma_r$ is 
taken to be surface at constant 
radial direction $r$. $^9K$ is the extrinsic curvature of the boundary,
\begin{align}
	^9 K = \nabla_\mu n^\mu = \frac{1}{\sqrt{\bf{g}}} 
\partial_\mu \left( \sqrt{\bf{g}}\, n^\mu \right)=\frac{1}{4}e^{-\Phi/4} e^{-k}\left[
9 \Phi' +  8 (g' + h') +  4 k')\right]
 \end{align}
 where $n^\mu$~is the boundary outward normal vector, $n^\mu= \sqrt{g^{rr}}\delta^\mu _ r$. Using the equations of motion  ${\cal I_{\rm grav}}$ reduces  to a volume integral of a  total derivative, 
\begin{align}
	 {\cal I_{\rm grav}}& =   \frac{1}{32 \pi } \int_{\cal M} d^{10}x \sqrt{\bf{g}}\nabla_\mu \nabla^\mu \Phi =   \frac{1}{ 32\pi} \int_{\cal M} d^{10}x \partial_\mu(\sqrt{\bf{g}} \bf{g}^{\mu\nu} \partial_\nu \Phi)\nn \\
&= vol_8 \beta \frac{1}{32 \pi }\lim_{r\to\infty}\left(\frac{1}{8} 
e^{2( \Phi + g + h)} \Phi' \right).
\end{align}
Explicitly, the surface  term is 
\begin{align}
	{\cal I_{\rm surf}}& = -vol_8  \beta \frac{1}{
	8\pi}\lim_{r\to\infty}\left\{
\frac{1}{32} e^{2 (\Phi + g + h)}\left[9 \Phi' +    8 (g' + h') + 4 k'\right]\right\}.
 \end{align}
Thus,
\begin{align}\label{eq:I-nonreg-final}
	{\cal I}= &{\cal I_{\rm grav}}+ {\cal I_{\rm surf}}\nn\\
=& -\frac{vol_8 \beta }{256 \pi}\lim_{r\to\infty}\left\{ e^{2 (\Phi + g + h)}\left[
8( \Phi' + 
   g' + h') + 
   4 k'\right]\right\}.
   \end{align}
   Equation \eqref{eq:I-nonreg-final} gives the value of the on-shell action in terms of the asymptotic fields at infinity. It  typically contains divergences and has to be regularized.  One way of doing this is to subtract the action of a reference background. In our case  the natural choice is to subtract a supersymmetric background. We also require that both backgrounds induce the same metric at the boundary, $\Sigma_r$,
\begin{align}\label{eq:match1-bdy-free}
	e^{\frac{\Phi_{ns}}{2}} e^{ 2 g_{ns}}&=e^{\frac{\Phi_{su}}{2}} e^{ 2 g_{su}},\quad\ e^{\frac{\Phi_{ns}}{2}} e^{ 2 h_{ns}}=e^{\frac{\Phi_{su}}{2}} e^{ 2 h_{su}},\nn\\
	e^{\frac{\Phi_{ns}}{2}} e^{ 2 k_{ns}}&=e^{\frac{\Phi_{su}}{2}} e^{ 2 k_{su}},\quad e^{\frac{\Phi_{ns}}{2}}=e^{\frac{\Phi_{su}}{2}} 
\end{align}
and that  the matter fields coincide at the boundary. 
In order to achieve the matching of the induced metrics and matter fields 
at the boundary we have to choose particular values for the integration constants 
of the supersymmetric background that we use as a regulator. We can 
then evaluate the free energy, 
\begin{align}
	F&=\frac{1}{\beta}( {\cal I}^{ns} -{\cal I}^{su} )\nn\\
	&= -\frac{vol_8  }{256 \pi}\lim_{r_c\to\infty} \left\{ e^{2 \Phi_{ns} + 2 g_{ns} + 2 h_{ns}}(8 \Phi_{ns}' + 
		8 g_{ns}' + 8 h_{ns}' + 
		4 k_{ns}')\right .\nn\\
		&\qquad\qquad\qquad\qquad \left .- e^{2 \Phi_s + 2 g_s + 2 h_s }(8 \Phi_s' + 
  8 g_s' + 8 h_s' + 
   4 k_s')  \right\}.
\end{align}
Using the UV expansion \eqref{UVexpansionnonsusy}, to first order in $W_{20}$,
\be
F = E= \frac{vol_8}{24 \pi} c^2_{+} e^{2\rho_0 + 2\Phi_\infty} W_{20}.
\ee
which agrees with the ADM calculation.
A similar evaluation of the free energy can be carried out for the backgrounds after the rotation. Due to the presence of  $F_5$ and the Chern-Simons term the calculation is more involved and the equality of the energy before and after rotation cannot be expressed as simply as \eqref{eq:beforeplusextra}. Nevertheless, after plugging in the appropiate UV expansions we  get, to first order in $W_{20}$,  $F_{before}\sim F_{after}\sim c^2_{+} e^{2\rho_0 + 2\Phi_\infty} W_{20}$ as expected.

\section{Appendix: Calculation of $B_2$}
\setcounter{equation}{0}
\label{B2appendix}
In the SUSY case, we have
\begin{align}
B_2	= \k \frac{e^{\frac{3}{2}\Phi}}{\hat{h}^{1/2}}\Big[e^{\r 3}-\cos\alpha
(e^{\theta\varphi}+ e^{12})-\sin\alpha(e^{\theta2}+ e^{\varphi 1})   \Big], \label{eq:SUSYB2}
\end{align}
with
\begin{align}
\cos\alpha=\frac{\cosh(2\r)-a}{\sinh(2\r)},\qquad
\sin\alpha=- \frac{2e^{h-g}}{\sinh(2\r)}.					\label{eq:sincosalpha}
\end{align}
This is not valid in the general non-SUSY case. We obtain the same $H_3$ as in the SUSY case \eqref{configurationfinal}, but the relationship to \eqref{eq:SUSYB2} requires the BPS equations, as does the consistency of the definitions \eqref{eq:sincosalpha}.

Instead, we must determine $B_2$ by requiring that $dB_2=H_3$. We assume that $B_2$ has the same general structure as \eqref{eq:SUSYB2},
\begin{align}
	B_2	=	b_1(\r) e^{\r 3} + b_2(\r) e^{\theta\varphi} + b_3(\r)e^{12} + b_4(\r)e^{\theta2} + b_5(\r)e^{\varphi 1},
\end{align}
which results in
\begin{align}
dB_2 &= \frac{e^{-h-k-\frac{\Phi }{4}} \left(a b_3 e^g+2 b_4 e^h\right)}{\hat{h}^{1/4}}   e^{1\theta3}    
				+  \frac{e^{-h-k-\frac{\Phi }{4}} \left(a b_3 e^g+2 b_5 e^h\right)}{\hat{h}^{1/4}}   e^{\varphi 23}   \nn\\ 
&\quad  -\frac{\left(b_4-b_5\right) e^{-h-\frac{\Phi }{4}} \cot \theta }{\hat{h}^{1/4}}   e^{\theta\varphi1}   \nn\\ 
&\quad  +  \frac{e^{-2 g-k-\frac{\Phi }{4}}}{2 \hat{h}^{5/4}} \left(e^{2 g} \left\{\hat{h} \left[b_3 \left(4 g'+\Phi '\right)+2 b_3'\right]+b_3 \hat{h}'\right\}+4 b_1 \hat{h} e^{2 k}\right)   e^{\rho12}   \nn\\ 
&\quad  +  \frac{e^{-h-k-\frac{\Phi }{4}}}{2 \hat{h}^{5/4}} \biggl(b_3 e^g \hat{h} a'-2 a b_1 e^{-g} \hat{h} e^{2 k}   \nn\\ 
&\quad\qquad\qquad\qquad  +e^h \left\{\hat{h} \left[b_4 \left(2 g'+2 h'+\Phi '\right)+2 b_4'\right]+b_4 \hat{h}'\right\}  \biggr)   e^{\rho\theta2}   \nn\\ 
&\quad  +  \frac{e^{-h-k-\frac{\Phi }{4}}}{2 \hat{h}^{5/4}} \biggl(b_3 e^g \hat{h} a'-2 a b_1 e^{-g} \hat{h} e^{2 k}   \nn\\ 
&\quad\qquad\qquad\qquad  +e^h \left\{\hat{h} \left[b_5 \left(2 g'+2 h'+\Phi '\right)+2 b_5'\right]+b_5 \hat{h}'\right\}  \biggr)    e^{\rho\varphi1}   \nn\\ 
&\quad  +  \frac{e^{- h-k-\frac{\Phi }{4}}}{2 \hat{h}^{5/4}} \biggl( -(b_4+b_5) e^g \hat{h} a'-\left(a^2-1\right) b_1 \hat{h} e^{2 k-h}   \nn\\ 
&\quad\qquad\qquad\qquad  +e^h \left\{\hat{h} \left[b_2 \left(4 h'+\Phi '\right)+2 b_2'\right]+b_2 \hat{h}'\right\}\biggr)   e^{\rho\theta\varphi }.
\label{eq:dB2}
\end{align}
Comparing with \eqref{configurationfinal}, we see that the $e^{\theta\varphi1}$ component of $H_3$ is zero, from which we immediately obtain that	$b_4 = b_5$. The $e^{\r\theta2}$ and $e^{\r\varphi1}$ components of (\ref{eq:dB2}) are then 
identical, as are the $e^{1\theta3}$ and $e^{\varphi23}$ components. This is also the case in $H_3$, so we are left with four independent equations.

Equating the $(e^{1\theta3}+e^{\varphi23})$ components results in
\begin{align}
	b_4 =			-\frac12 e^{g-h} a b_3		-		\frac{	\k N_c e^{\frac{3\Phi}{2}-g-h} b'	 }
																							 {4\hat{h}^{1/2}}	,
\end{align}
and the $e^{\r12}$ component gives
\begin{align}
	b_1	&=		\frac{e^{2g-2k}}{4\hat{h}} \Bigl[		2b_3 \Phi' - 3\hat{h}b_3 \Phi'  - 4\hat{h}b_3 g' - 2\hat{h}b_3'			\nn\\
			&\qquad\qquad\qquad\qquad\qquad\qquad			+\k N_c e^{\frac{3\Phi}{2}-2h} \hat{h}^{\frac12} \left(	a^2 - 2ab +1	\right)
																				\Bigr].
\end{align}
This leaves $b_2$ and $b_3$ to be determined. Substituting these results into \eqref{eq:dB2}, we find that the $(e^{\r\theta2}+e^{\r\varphi1})$ component of $H_3=dB_2$ reduces to the equation of motion \eqref{eq:eomb} for $b$. The only remaining equation is then the $e^{\r\theta\varphi}$ component. This is a first order differential equation in $b_2$ and $b_3$,
\begin{align}
0&=				8 \hat{h}  e^{2 g+4 h} b_2'
					+2 \left(a^2-1\right) \hat{h}  e^{4 g+2 h}	 b_3'
					+e^{2 (g+h)} \hat{h}' \left[\left(a^2-1\right) e^{2 g}b_3 + 4  e^{2 h}b_2\right]		\nn\\
 &\qquad      +\hat{h} e^{2 (g+h)} \left[4 a e^{2 g} a'b_3 + (a^2-1) e^{2 g} \left(4 g'+\Phi '\right)b_3 +4 b_2 e^{2 h} \left(4 h'+\Phi '\right)\right]			\nn\\
 &\qquad      -\kappa N_c \sqrt{\hat{h}}   e^{3 \Phi /2} \Bigl[-2 a' b' e^{2 (g+h)}+(a^4-1) e^{4 g}		\nn\\
 &\qquad\qquad\qquad\qquad\qquad\qquad      -2 (a^2-1) ab e^{4 g}+2 a b e^{4 g}-16 e^{4 h}\Bigr]    .   \label{eq:b2b3diff}
\end{align}
Solving for $b_2$ we obtain
\begin{align}
b_2 &= \frac{e^{-2h-\Phi/2}}{\sqrt{\hat{h}}} \int^\rho d\rho' \Biggl(
\frac{e^{-2 g-2 h+\frac{\Phi }{2}}}{8 \sqrt{\hat{h}}} \biggl\{-\left(a^2-1\right)  e^{4 g+2 h} \hat{h}'	b_3	\nn\\
		&\quad			- \hat{h} e^{4 g+2 h} \Bigl[4 a a'+a^2 \left(4 g'+\Phi '\right)-4 g'-\Phi '\Bigr] b_3		\nn\\
		&\quad			+\k N_c  \sqrt{\hat{h}} e^{3 \Phi /2} \left[(a^4-1) e^{4 g}-2 (a^2-1)a b e^{4 g}-2 a' b' e^{2 (g+h)}-16 e^{4h}\right]\biggr\}		\nn\\
		&\quad			-\frac{1}{4} \left(a^2-1\right) \sqrt{\hat{h}}  e^{2 g+\frac{\Phi }{2}}b_3'
\Biggr),
\end{align}
which does not appear to be very useful. Instead, we can use the fact that we want $Q_{\text{\text{Page, D3}}}=0$ (see eq. \ref{eq:chargedefs}). We therefore impose that the $e^{\theta\varphi123}$ component of $F_5 - B_2\wedge F_3$ vanishes. The resulting equation is algebraic in $b_2$ and $b_3$, and results in
\begin{align}
	b_2 &=		\frac{e^{-2h}}{4\hat{h}^{1/2}} \left\{	e^{2g}\hat{h}^{\frac12} \left(  1-a^2  \right)b_3
																											-\frac{\k}{N_c} e^{\frac{3\Phi}{2}} \left[	N_c^2 (a-b) b' + 4e^{2(g+h)} \Phi'
																																							\right] 
																							\right\}.	
\end{align}
Together with the above results for $b_{1,4,5}$ this completes \eqref{eq:bs}.

It remains to check that this $b_2$ is also compatible with the requirement that $dB_2=H_3$. Substituting into \eqref{eq:b2b3diff} we find that $b_3$ cancels, giving
\begin{align}
0	&=	4 e^{4 (g+h)} \left\{2 \hat{h} \left[2 g' \Phi '+2 h' \Phi '+\Phi ''+2 \left(\Phi '\right)^2\right]-2 g' \hat{h}'-2 h' \hat{h}'-\hat{h}''\right\}	\nn\\
	&\qquad\quad			+N_c^2 \Bigl[a^4 e^{4 g}-2 a^3 b e^{4 g}+2 (a-b) b'' e^{2 (g+h)}+4 (a-b) b' e^{2 (g+h)} \Phi '		\nn\\
	&\qquad\quad			+2 a b e^{4 g}-2 \left(b'\right)^2 e^{2 (g+h)}-e^{4 g}-16 e^{4 h}\Bigr]	.
\end{align}
This is solved by the equations of motion (\ref{eq:eomphi}, \ref{eq:eomb}) for $\Phi$ and $b$.

To determine the effect of the undetermined function $b_3$, we can look at the difference $\Delta B_2 = B_2 - (B_2)_{b_3=0}$, which we find to be 
of the form
\begin{align}
	\Delta B_2 &= F_1(\rho) \sin\theta\ d\theta\wedge d\varphi
							+F_2(\rho) \sin\tilde{\theta}\ d\tilde{\theta}\wedge d\tilde{\varphi}
							+F_3(\rho) \cos\theta\ d\r\wedge d\varphi			\nn\\
	&\qquad\qquad\qquad\qquad
							+F_4(\rho) \cos\tilde\theta\ d\r\wedge d\tilde\varphi
							+F_5(\rho) \ d\r\wedge d\psi ,
\end{align}
where the $F_i$ depend on $g$, $\Phi$, $\hat{h}$, $b_3$ and their derivatives. If we set this equal to
\begin{align}
	d\left[ \beta_1(\rho)\cos\theta\ d\varphi
					+ \beta_2(\rho)\cos\tilde\theta\ d\tilde\varphi
					+ \beta_3(\rho)\ d\psi
	 \right]
\end{align}
we can solve for the $\beta_i$, giving
\begin{align}
	\Delta B_2 &= -\frac14 d\left[	e^{2g+\Phi/2} \sqrt{\hat{h}} b_3 (\cos\theta\ d\varphi
																																		+\cos\tilde\theta\ d\tilde\varphi
																																		+d\psi
																																		)				\right]			\nn	\\
						&= -\frac{1}{2} d\!\left(  e^{2g-k+\Phi/4} \hat{h}^{1/4} b_3 \  e^3     \right).
\end{align}

\section{Appendix: Seiberg-like duality}
\setcounter{equation}{0}
\label{seibergappendix}
In section \ref{seibergsection} we discuss how the operation known as Seiberg duality 
in the KS cascade acts for our non-SUSY solution.
In order to do so, we find it instructive to compare to two
different cases: the KS case and the baryonic branch case. These are summarized here.

\subsection{The KS case}
We follow here the treatment
in 
\cite{Benini:2007gx}, 
specified in the case of no flavors ($N_f=0$).
The NS potential $B_2$ is given by,
\beq
B_2=\frac{N_c}{2}[f g_1\wedge g_2 + \tilde{k} g_3\wedge g_4]
\eeq
where the definition of $g_1,....,g_4$ can be found in \cite{Benini:2007gx}.
When specialized to the cycle
\beq
\Sigma_2=[\theta=\tilde{\theta}, \varphi=2\pi-\tilde{\varphi}, \psi=\psi_0 ]
\eeq
we obtain that
\beq
B_2|_{\Sigma_2}=\frac{N_c}{2}[(f+\tilde{k}) +(\tilde{k}-f)\cos\psi_0    ]\sin\theta d\theta \wedge d\varphi
\eeq
from which one finds
\beq
b_0=\frac{1}{4\pi^2}\int_{\Sigma_2}B_2=\frac{N_c}{\pi}
[f\sin^2(\frac{\psi_0}{2}) + k\cos^2(\frac{\psi_0}{2})]
\eeq
On the other hand, as computed in
\cite{Benini:2007gx}, we can see that the Maxwell charge of D3 branes is
\beq
Q_{Max, D3}=\frac{N_c^2}{\pi}[f-(f-\tilde{k})F]
\eeq
We see that under the change
\beq
f\to f-\frac{\pi}{N_c},\;\;\; 
\tilde{k}\to \tilde{k}-\frac{\pi}{N_c}
\eeq
the D3-Maxwell charge changes by 
\beq
Q_{Max, D3}\to Q_{Max, D3}-N_c,
\;\;\; 
b_0\to b_0-1.
\label{changes}
\eeq
these transformations, are equivalent to
changing the NS potential with a large gauge transformation
\beq
B_2\to B_2 +\frac{\pi}{2}[g_1\wedge g_2 + g_3\wedge g_4]
\eeq
which when evaluated on the cycle $\Sigma_2$, produces the changes in 
eq.(\ref{changes}).
We move now to analyze the baryonic branch (SUSY) case.
\subsection{Baryonic branch case}
In this case the NS potential is
\beq
B_2=\frac{\k e^{3\Phi/2}}{\hat{h}^{1/2}}[e^{\r3}-\cos\a 
(e^{12}+e^{\theta\varphi}) -\sin\a (e^{\theta2} +e^{\varphi 1})]
\eeq
Evaluating this on the $\Sigma_2$ we get
\bea
& & b_0=\frac{\k e^{2\Phi}}{\pi}\Big[ (\tilde{k}+f)+(\tilde{k}-f)\cos\psi_0    \Big],\nonumber\\
& & \tilde{k}+f =\frac{\k e^{2\Phi}}{N_c}
\Big[ \cos\a(\frac{e^{2g}}{4}(a^2+1)-e^{2h}          ) +\sin\a a e^{h+g} \Big],\nonumber\\
& & \tilde{k}-f =\frac{\k e^{2\Phi}}{N_c}
\Big[ \cos\a\frac{e^{2g}}{2}a  +\sin\a  e^{h+g} \Big],
\eea
Using the explicit expressions, we have
\beq
\tilde{k}=-\frac{\k e^{2\Phi}}{4N_c}Q \coth(\r), \;\;\;
f=-\frac{\k e^{2\Phi}}{4N_c}Q \tanh(\r)
\eeq
The Maxwell charge for D3 branes can be written as,
\beq
Q_{Max, D3}=\frac{\k}{\pi} e^{2g+2h+2\Phi}\Phi'
\eeq
and using the BPS equation for $\Phi'$
we have
\bea
& & Q_{Max, D3}=\frac{N_c^2}{\pi}\Big[  2f+ (\tilde{k}-f)F \Big],
\nonumber\\
\eea
where  $F=(1-b)$.
So, once again, we obtain that 
under a large gauge transformation, 
\beq
b_0\to b_0-1,\;\;\; Q_{Max, D3}\to Q_{Max, D3}- N_c.
\eeq

\end{document}